\documentclass[11pt,a4paper]{article}
\pdfoutput=1

\usepackage{jheppub}
\usepackage{rotating}

\usepackage{wrapfig}
\usepackage{amsfonts}
\usepackage{amsmath}
\usepackage{slashed}
\usepackage{amssymb}
\usepackage{latexsym}
\usepackage{multirow}
\usepackage{hyperref}
\usepackage{enumitem}
\hypersetup{colorlinks=true,citecolor=red,linkcolor=magenta,urlcolor=blue}
\usepackage{subfigure}
\usepackage{relsize}
\usepackage{booktabs}
\usepackage{cleveref}
\usepackage{fancyhdr}
\usepackage{float}
\usepackage{graphicx}
\usepackage{microtype}
\usepackage{tabularx}
\usepackage{xcolor}
\usepackage[bottom]{footmisc}
\usepackage{cancel}

\title{On the BSM reach of four top production at the LHC}
\abstract{
Many scenarios of beyond the Standard Model (BSM) physics give rise to new
top-philic interactions that can be probed at proton machines such as the Large
Hadron Collider through a variety of production and decay modes. On the one
hand, this will enable a detailed determination of the BSM model's parameters
when a discovery is made and additional sensitivity in non-dominant production modes
can be achieved. On the other hand, the naive narrow width approximation in dominant
production modes such as gluon fusion might be inadequate for some BSM parameter
regions due to interference effects, effectively making less dominant production modes
more relevant in such instances. In this work, we consider both these questions
in the context of four top quark final states at the LHC. Firstly, we show that the SM 
potential can be enhanced through the application of targeted Graph Neural Network techniques
that exploit data correlations beyond cut-and-count approaches. Secondly, we show that
destructive interference effects that can degrade BSM sensitivity of top-philic states from 
gluon fusion are largely avoided by turning to four top final states. This achieves considerable exclusion potential
for, e.g., the two Higgs doublet model. This further motivates
four top final states as sensitive tools for BSM discovery in the near future of the LHC. 
}
\author[a]{Anisha,}  
\author[a]{Oliver Atkinson,}  
\author[a]{Akanksha Bhardwaj,}
\author[a]{Christoph Englert,} 
\author[a]{Wrishik Naskar,}
\author[b]{Panagiotis Stylianou}
\affiliation[a]{School of Physics \& Astronomy, University of Glasgow, Glasgow G12 8QQ, United Kingdom}
\affiliation[b]{Deutsches Elektronen-Synchrotron DESY, Notkestr. 85, 22607 Hamburg, Germany}
\emailAdd{anisha@glasgow.ac.uk} 
\emailAdd{o.atkinson.1@research.gla.ac.uk} 
\emailAdd{akanksha.bhardwaj@glasgow.ac.uk}
\emailAdd{christoph.englert@glasgow.ac.uk}
\emailAdd{w.naskar.1@research.gla.ac.uk}
\emailAdd{panagiotis.stylianou@desy.de}

\preprint{DESY-23-025}

\begin{document}
\maketitle
\flushbottom
%%%%%%%%%%%%%%%%%%%%%%%%%%%%%%%%%%
\section{Introduction}
\label{sec:intro}
%%%%%%%%%%%%%%%%%%%%%%%%%%%%%%%%%%
The absence of evidence for new physics beyond the Standard Model (BSM) in searches at the high-energy
frontier of the Large Hadron Collider (LHC) is on the one hand puzzling, as the fundamental origin of
the electroweak scale and its stability remains a mystery. On the other hand, the BSM benchmarking programme
leading up to the LHC data was, perhaps, overly optimistic.
The absence of a BSM signal has highlighted less abundant collider processes as important tools for BSM discrimination,
drawing from model-independent and theoretically motivated techniques as well as improved approaches to better exploiting
model-(un)specific correlations in searches for new interactions. The resurgence of machine learning techniques 
applied to the LHC realm is an example of the latter; effective field theory (EFT) in its linear or non-linear formulation is 
an example of the former.

Particularly interesting processes along these lines are four top quark final states, with the ATLAS and CMS collaborations recently reporting evidence of SM-like production~\cite{CMS:2019rvj,ATLAS:2020hpj,ATLAS:2021kqb,CMS:2022uga,CMS:2019jsc,ATLAS:2018kxv}. Four top final states are relevant processes when considering EFT deformations of expected SM correlations~\cite{Zhang:2017mls,Banelli:2020iau,Aoude:2022deh}, lending considerable sensitivity to large-scale EFT fits, but they also enable a targeted snapshot of electroweak Higgs properties~\cite{Englert:2019zmt}. Recently in~\cite{Builtjes:2022usj}, it was demonstrated that the large kinematical information that can be exploited in such a busy final state leads to improved sensitivity when final state correlations are tensioned against the SM backgrounds using machine learning approaches. 

For top-philic resonance extensions, four top final states are naively less motivated as gluon fusion provides a large production cross section and top pair resonance searches are experimentally less challenging than the analysis of four top states. Given the generic vulnerability of $pp\to t\bar t$
final states to quantum interference with QCD backgrounds~\cite{Gaemers:1984sj,Dicus:1994bm}, this might well be a fallacy: when BSM signals are rendered small through interference
in the dominant production and decay channels, only those with limited production cross sections remain. Higgs pair production as a relatively clean tool
in such circumstances can be statistically limited and four top final states might well provide a more robust alternative route for discovery.

The aim of this work is to clarify both these avenues: building on sensitivity enhancements in discriminating SM four top production, we quantitatively
analyse the sensitivity improvement when turning to motivated extensions of the SM. A further emphasis is given to the relevance of interference 
effects in four top final states.

This paper is organised as follows. In Sec.~\ref{sec:analysis}, we provide details on event simulation and inclusive fiducial selection for the different
(lepton-dominated) partonic collider processes that form the basis of this study: same-sign dilepton in association with $b$ quark and light jets
\begin{subequations}
\begin{alignat}{3}
p p \to t \bar t t \bar t \to \ell^+ \ell^+ / \ell^- \ell^- + \text{jets} + \text{$b$ quarks}&\quad\hbox{(2SSDL)},
\end{alignat}
and three lepton production in association with $b$ and light flavour jets 
\begin{alignat}{3}
p p \to t \bar t t \bar t \to \ell \ell \ell + \text{jets} +\text{$b$ quarks}&\quad\hbox{(3L)}.
\end{alignat} 
\end{subequations}
In the following, $\ell = e, \mu$ and we include leptonic decays of $\tau$ leptons (see~\cite{CMS:2019rvj,ATLAS:2021kqb} for recent explorations by the ATLAS 
and CMS experiments).
Subsequently, we detail in Sec.~\ref{sec:analysis} our analysis implementation in terms of a Graph Neural Network (GNN) that
provides a particularly motivated approach to exploiting the various particle-level correlations expected in signal and background
distributions. Sec.~\ref{sec:results} is devoted to results and we first discuss the sensitivity yield for SM four top production (Sec.~\ref{sec:smsig}), which provides
the baseline for the subsequent BSM sensitivity discussion in Secs.~\ref{sec:nonres} and~\ref{sec:res}. Our BSM discussion is split into addressing the relevance
of GNN techniques when facing resonant and non-resonant BSM extensions to gauge the potential of the approach for different phenomenological
situations. For resonant BSM extensions, we also touch on the relevance of signal-background interference, which is known to be large in resonant di-top final states~\cite{Jung:2015gta,Frederix:2007gi,Carena:2016npr,Hespel:2016qaf,Djouadi:2019cbm,ATLAS:2017snw,CMS:2019pzc}, but turns out to be negligible for the
dominant four top production modes. A summary and conclusions are provided in Sec.~\ref{sec:conc}.

%%%%%%%%%%%%%%%%%%%%%%%%%%%%%%%%%%
\section{Analysis Framework}
\label{sec:analysis}
%%%%%%%%%%%%%%%%%%%%%%%%%%%%%%%%%%
\subsection{Event simulation and fiducial selection}
To prepare our datasets we need to preselect the events appropriate for 2SSDL 
and 3L
final state topologies. Events are generated from simulating proton-proton collisions at $\sqrt{s}=13\;\text{TeV}$ using \textsc{MadGraph}\_aMC@NLO~\cite{Alwall:2014hca} with leading order precision. These events are subsequently showered and hadronised using \textsc{Pythia} 8.3~\cite{Bierlich:2022pfr}. We then reconstruct the final states particles using~\textsc{MadAnalysis}~\cite{Conte:2012fm} that interfaces {\sc{FastJet}}~\cite{Cacciari:2011ma,Cacciari:2005hq}.

%%%%%%%%%%%%%%%%%%%%%%%%%%%%%%%%%%
\begin{table}[!t]
\centering
\begin{tabular}{|l|c|c|}
\hline\hline
Processes              &Cross Section (fb) \\ \hline\hline
$pp\rightarrow t_{\ell^+}\bar{t}_{h}W^+_{\ell^+}+t_{h}\bar{t}_{l_-}W^-_{\ell^-}$ & $57.67\pm$0.06  \\[0.1cm]
$pp\rightarrow t_{\ell^+}\bar{t}_{h}Z_{\ell^+\ell^-}+t_{h}\bar{t}_{\ell^-}Z_{\ell^+\ell^-}+t_{\ell^+}\bar{t}_{\ell^-}Z_{\ell^+\ell^-}$ & $10.65\pm0.01$  \\[0.1cm]
$pp\rightarrow (W^+_{\ell^+}W^-_{h}W^+_{\ell^+} + W^+_{h}W^-_{\ell^-}W^-_{\ell^-})b \bar{b}$ & $43.29\pm0.05$   \\[0.1cm]
$pp\rightarrow (W^+_{\ell^+}W^-_{h}Z_{\ell^+\ell^-}+W^+_{h}W^-_{\ell^-}Z_{\ell^+\ell^-}+W^+_{\ell^+}W^-_{\ell^-}Z_{\ell^+\ell^-}) b \bar{b}$ & $12.65\pm0.02$   \\[0.1cm]
\hline\hline
\end{tabular}
	\caption{\label{tab:2ssdl_prod}Cross sections for the different SM processes having the most significant contributions to the 2SSDL background, in accordance with our baseline cuts. The subscript on each particle describes the particle's decay channel ($\ell^\pm$ describes leptonic decay and $h$ refers to a hadronic decay channel). We include the uncertainties from the Monte-Carlo simulation.}
\end{table}
%%%%%%%%%%%%%%%%%%%%%%%%%%%%%%%%%%
%%%%%%%%%%%%%%%%%%%%%%%%%%%%%%%%%%
\begin{table}[!b]
\centering
\begin{tabular}{|l|c|c|}
\hline\hline
Processes              & Cross Section (fb)\\ \hline\hline
$pp\rightarrow t_{\ell^+}\bar{t}_{\ell^-}W^\pm_{\ell^\pm}$ & $3.421\pm0.004$ \\[0.1cm]
$pp\rightarrow t_{\ell^+}\bar{t}_{h}Z_{\ell^+\ell^-}+t_{h}\bar{t}_{\ell^-}Z_{\ell^+\ell^-}+t_{\ell^+}\bar{t}_{\ell^-}Z_{\ell^+\ell^-}$ & $10.65\pm0.01$ \\[0.1cm] 
$pp\rightarrow Z_{\ell^+\ell^-}W^\pm_{\ell^\pm} b \bar{b}$ & $3.296\pm0.003$ \\[0.1cm]
$pp\rightarrow W^+_{\ell^+}W^-_{\ell^-}W^\pm_{\ell^\pm} b \bar{b}$ & $3.614\pm0.004$ \\[0.1cm]
$pp\rightarrow (W^+_{\ell^+}W^-_{h}Z_{\ell^+\ell^-}+W^+_{h}W^-_{\ell^-}Z_{\ell^+\ell^-}+W^+_{\ell^+}W^-_{\ell^-}Z_{\ell^+\ell^-}) b \bar{b}$ & $12.65\pm0.02$  \\[0.1cm]
\hline\hline
\end{tabular}
	\caption{\label{tab:3l_prod} Summary of background cross sections and Monte-Carlo uncertainties for the different SM processes contributing to the 3L selection, given our baseline cuts. The subscript on each particle denotes the decay channel as in Tab.~\ref{tab:2ssdl_prod}.}
\end{table}
%%%%%%%%%%%%%%%%%%%%%%%%%%%%%%%%%%

To identify our inclusive search region we employ the following baseline selection criteria.
Light leptons (electrons and muons) are defined from a threshold of $p_T > 10~\text{GeV}$ within the detector coverage of the electromagnetic calorimeter $|\eta|<2.5$, where $p_T$ and $\eta$ denote the transverse momentum and pseudorapidity, respectively.
Selected $b$-tagged jets are identified from a threshold $p_T > 20~\text{GeV}$ within the tracker $|\eta|<2.5$. Light flavour jets that fall into the hadronic calorimeter coverage, $|\eta|<4.5$, are fed to the analysis if they satisfy $p_T > 20\;\text{GeV}$. Furthermore, we require a significant amount of missing transverse energy of at least $20~\text{GeV}$. To isolate the mis-measurement of missing transverse energy ($\vec{\slashed{E}_T}$) from jets we select the events with azimuthal angle difference $|\phi(j)- \phi{(\vec{\slashed{E}_T)}}| > 0.2$.

Our analysis for the two signal topologies 2SSDL 
and 3L
closely follows the ATLAS and CMS studies of Ref.~\cite{CMS:2019rvj,ATLAS:2020hpj,ATLAS:2021kqb,CMS:2022uga,CMS:2019jsc,ATLAS:2018kxv}. 
For the 2SSDL 
final state, 
we require at least 2 leptons, at least two $b$ jets and at least one additional jet according to the above criteria.
To select the events for the 3L 
search, we require at least three leptons, at least two $b$ jets without restriction on the light flavour jets.
Throughout, we employ a $70\%$ flat $b$-tagging efficiency, which corresponds to a pessimistic $b$ tagging working point~\cite{ATLAS:2012ima} (see also~\cite{Bols:2020bkb}). In turn this removes large contributions from mis-identified $b$ jets \cite{Bols:2020bkb} for the considered $p_T$ range of $b$ jets, hence we neglect effects from mis-identified $b$ jets  in this study. 
%%%%%%%%%%%%%%%%%%%%%%%%%%%%%%%%%%
\begin{figure}[!t]
\centering
\includegraphics[width=0.48\textwidth]{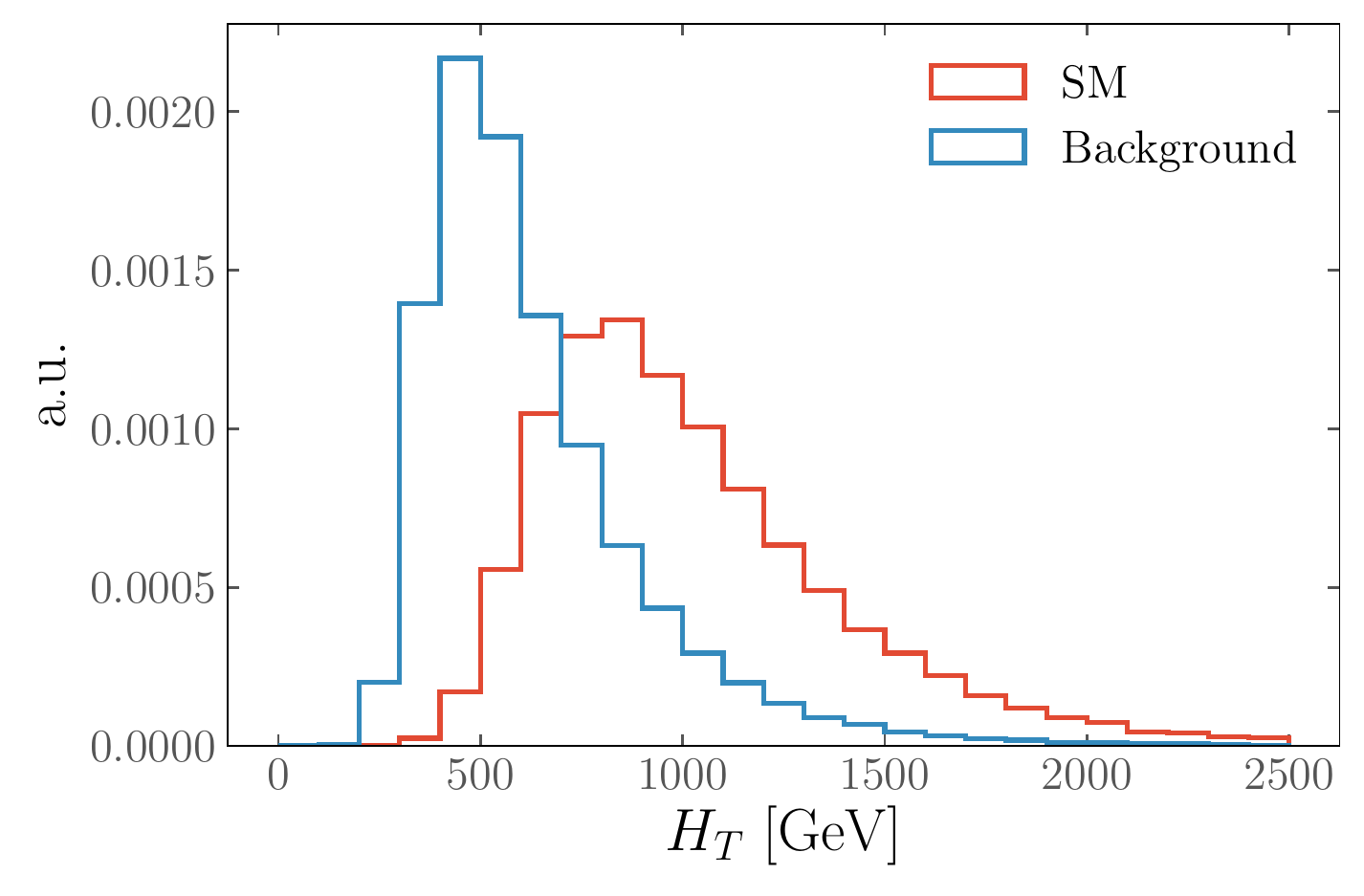}\hfill
\includegraphics[width=0.48\textwidth]{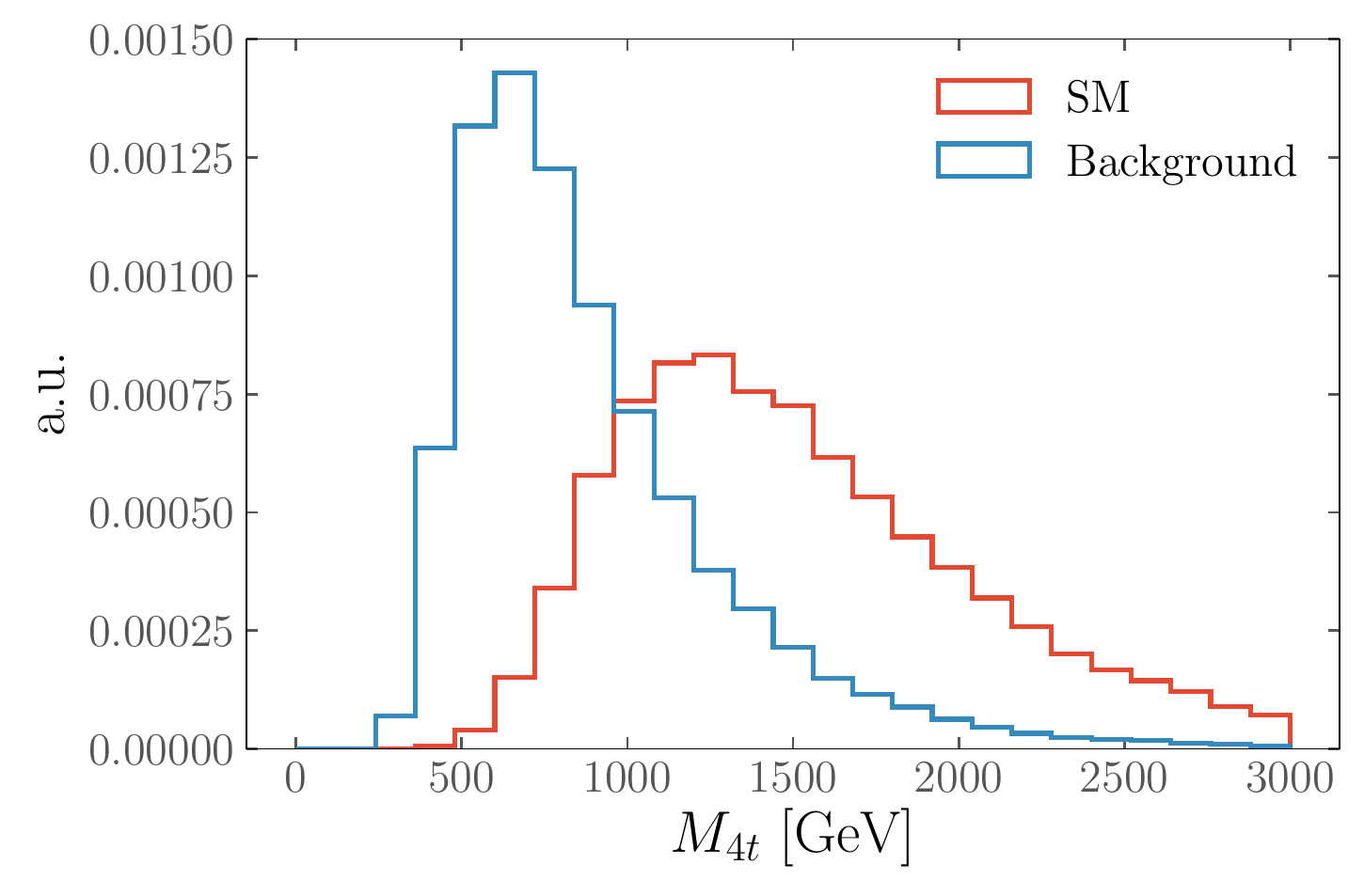}\\
\includegraphics[width=0.48\textwidth]{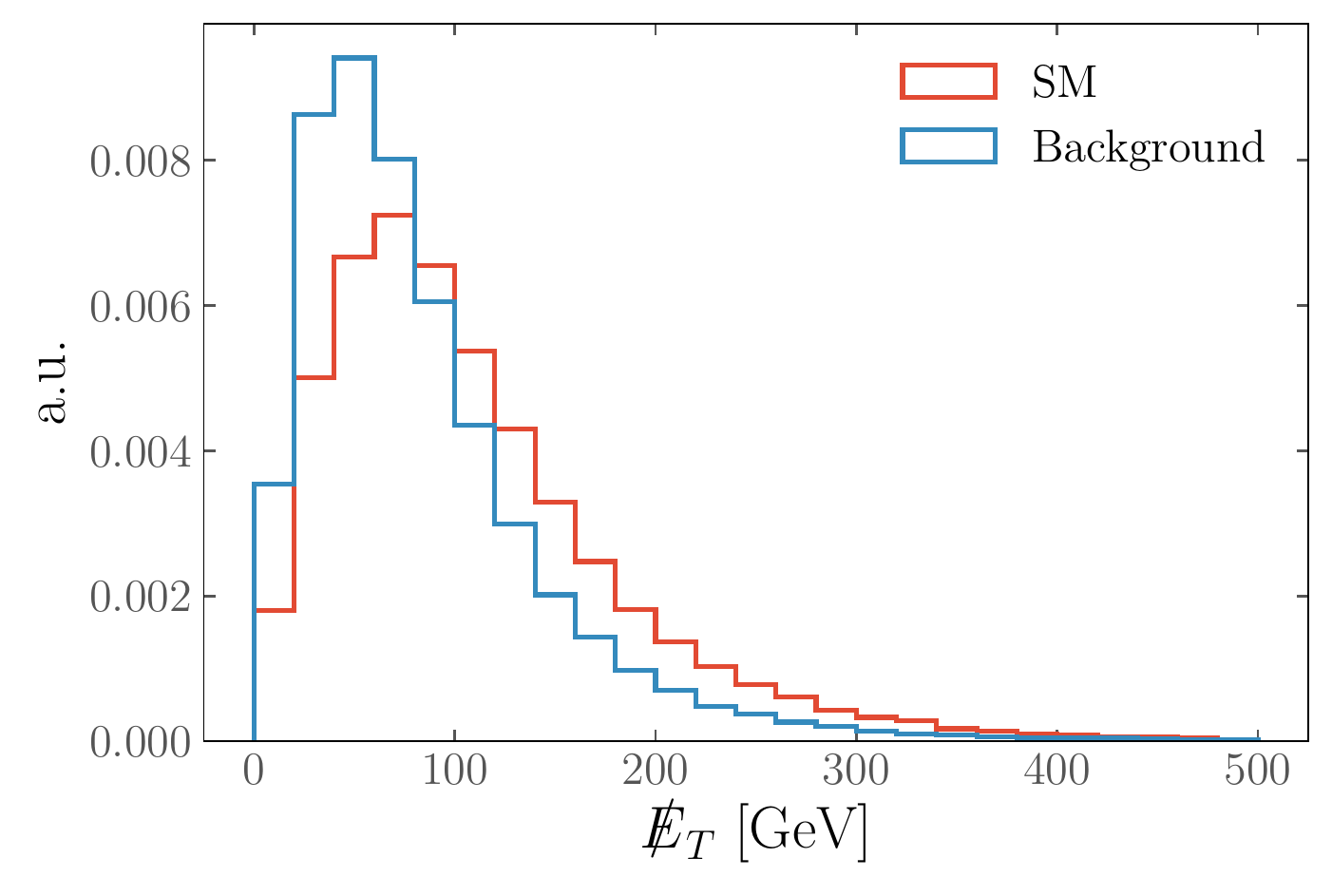}\hfill
\includegraphics[width=0.48\textwidth]{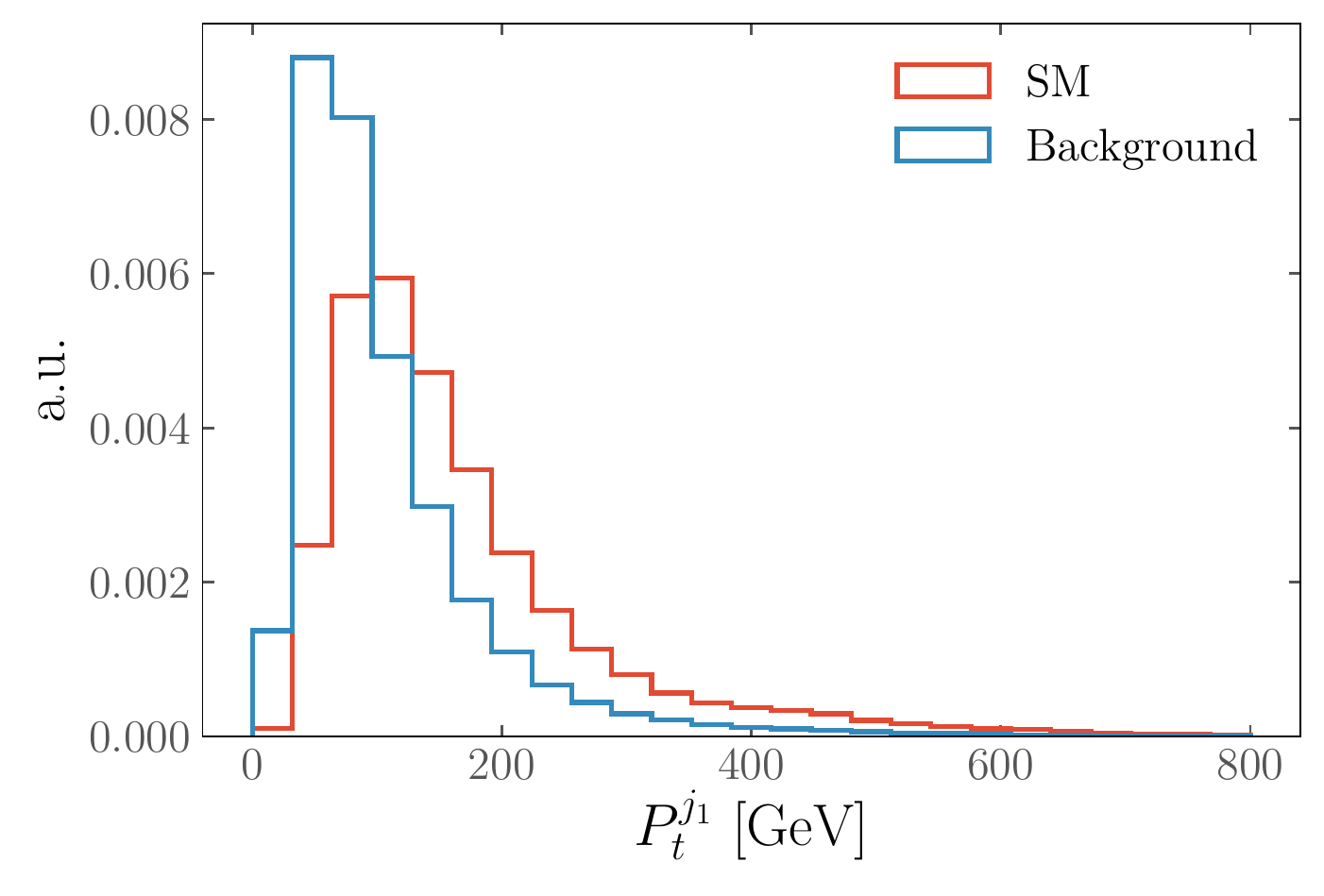}\\
\includegraphics[width=0.48\textwidth]{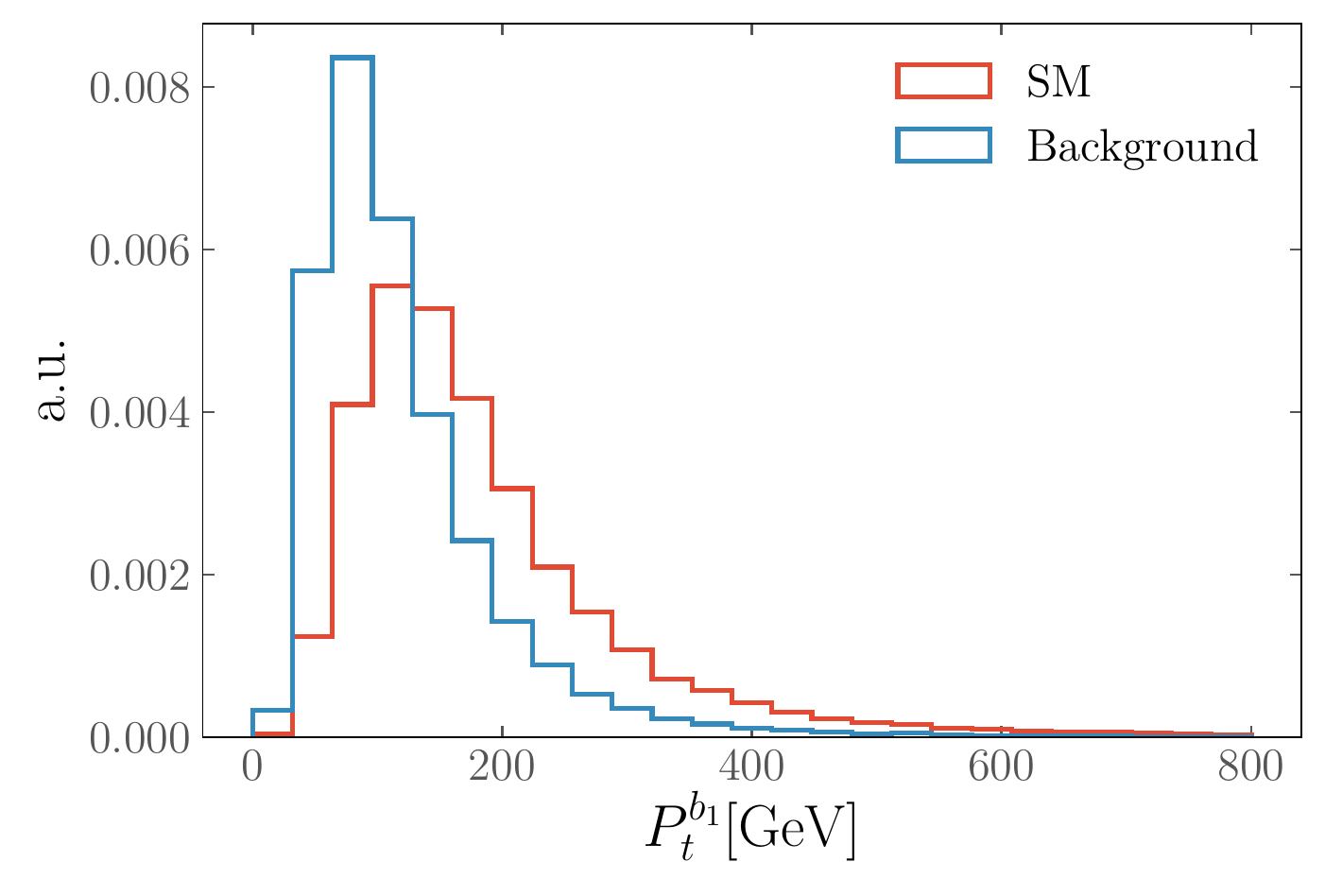}\hfill
\includegraphics[width=0.48\textwidth]{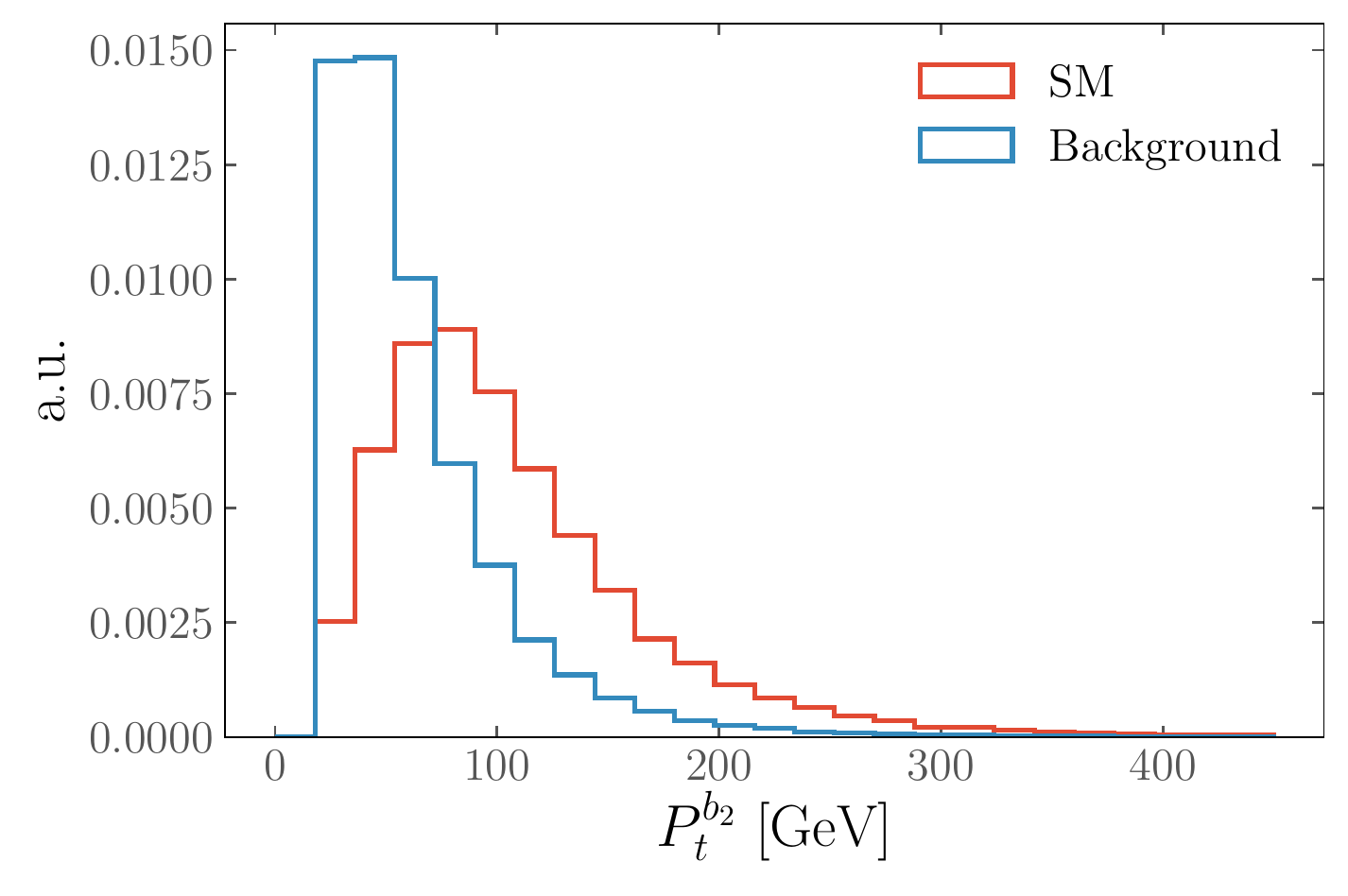}
\caption{Normalised distributions of the kinematic variables related to 2SSDL channel for SM four top signal and SM backgrounds contributing to the process at the LHC running at 13~TeV. \label{fig:KinematicDist}}
\end{figure}
%%%%%%%%%%%%%%%%%%%%%%%%%%%%%%%%%%

Having defined our fiducial regions, we turn to the dominant SM backgrounds contributing to these final states:
\begin{enumerate}
\item The $t\bar{t}W^\pm$ channel with leptonic decays of the tops and $W$ is the most dominant background for the 2SSDL 
case; however its contributions are subdominant for the 3L 
channel. 
\item Conversely, background contributions from $t\bar{t}Z$ are significant for the 3L case, while contamination of the 2SSDL signal region from this channel is reduced (yet substantial) compared to others.
\item $W^+W^-W^\pm+2b~\text{jets}$, with subsequent decays $W \to \ell \nu_\ell$, is the second large background for 2SSDL and remains a considerable background for the 3L selection. 

\item $W^+W^-Z+2b~\text{jets}$ with leptonic decays of vector bosons contributes as a background for both the channels, and yields the largest contamination in the 3L case. 
\item $ZW^\pm+2b~\text{jets}$ is a subdominant background to both channels when both $Z$ and $W$ bosons decay leptonically. However, its contribution to 2SSDL 
is negligible and therefore dropped from Tab.~\ref{tab:2ssdl_prod}. 
\end{enumerate}
The cross sections after decay and baseline selection are shown in Tabs.~\ref{tab:2ssdl_prod} and \ref{tab:3l_prod} contributing to  2SSDL 
and 3L 
channels, respectively.

%%%%%%%%%%%%%%%%%%%%%%%%%%%%%%%%%%
\begin{figure}[!t]
\centering
\includegraphics[width=0.48\textwidth]{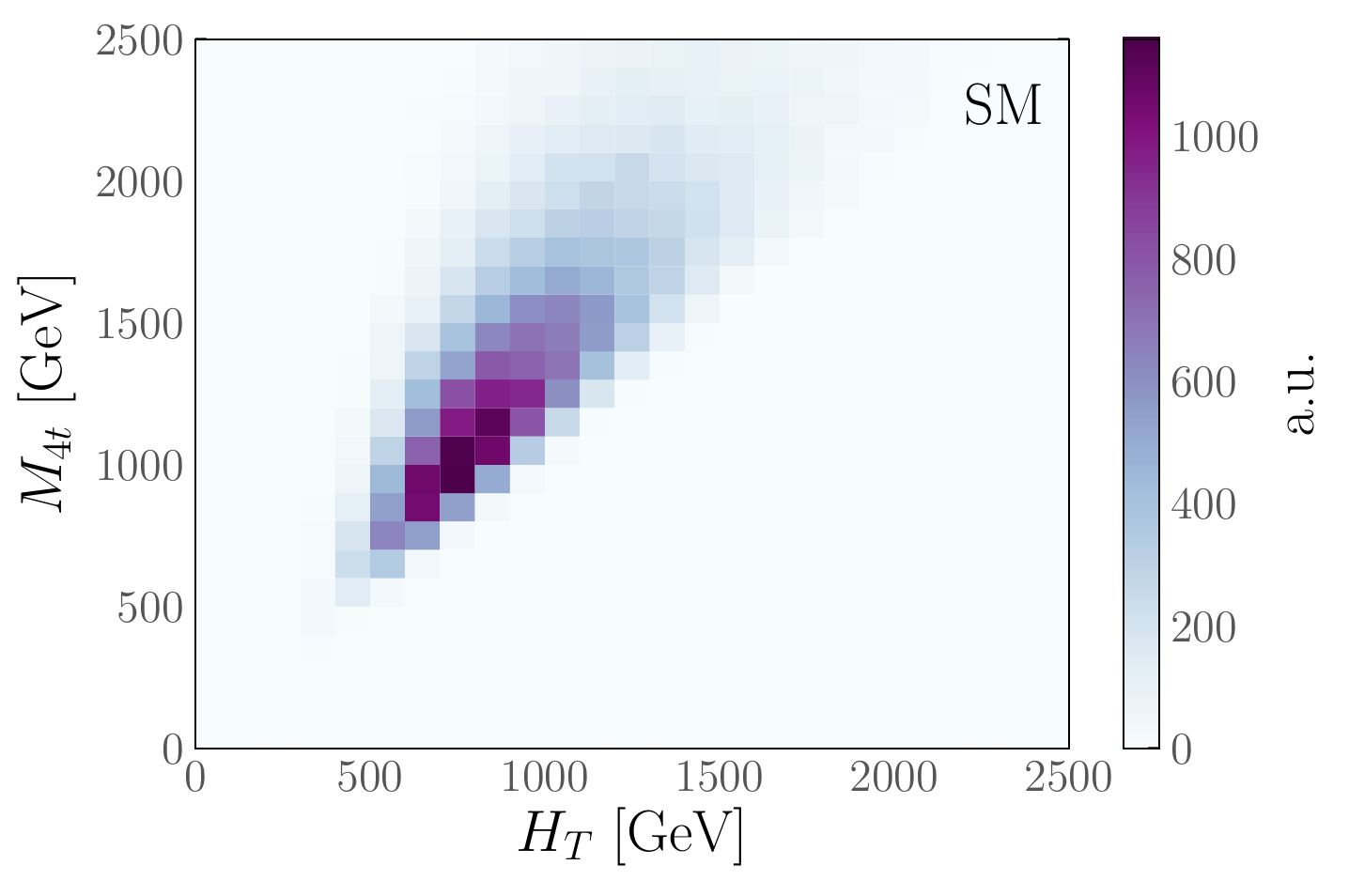}\hfill
\includegraphics[width=0.48\textwidth]{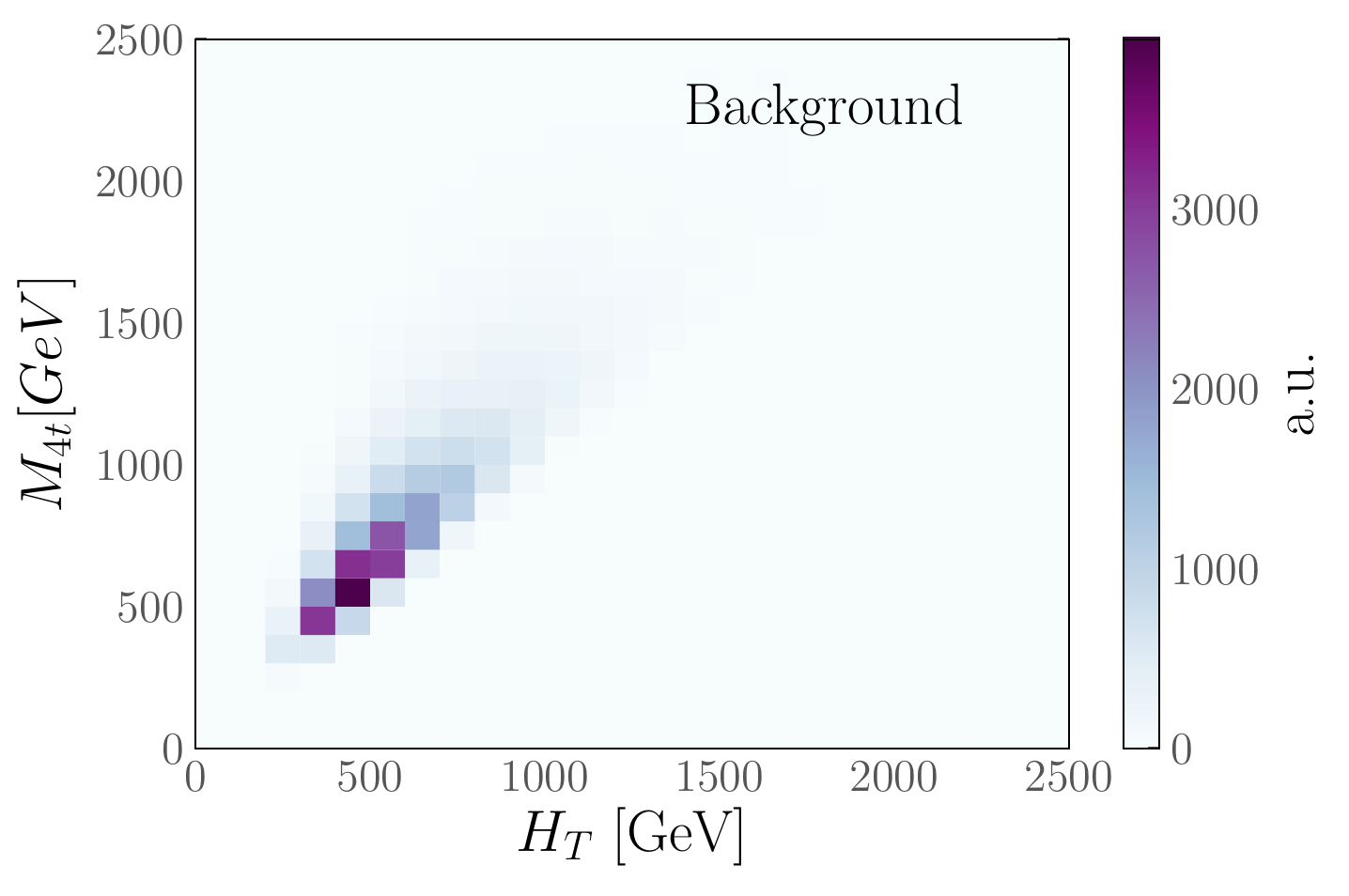}
\caption{Normalised two-dimensional distributions of the kinematic variables $H_T$ and $M_{4t}$ for 2SSDL channel for SM four tops signal and SM backgrounds contributing to the process at the LHC running at 13~TeV.
\label{fig:KinematicDist2D}}
\end{figure}
%%%%%%%%%%%%%%%%%%%%%%%%%%%%%%%%%%

We prepare data sets for the background class of our neural network setup by combining individual background samples according to their cross section to obtain a realistic composition expected at a fixed luminosity.
Representative kinematic distributions are shown in Fig.~\ref{fig:KinematicDist}, comparing signal and background at hadron level. It is evident that $H_T$ and $M_{4t}$\footnote{At hadron level we identify $M_{4t}$ as the total invariant mass of the visible final states of the system.} are the most discriminating observables and hardness-related observables can be expected as the dominant drivers of signal vs. background discrimination. Representative correlations that a neural network can exploit in a non-rectangular and optimised way are shown in Fig.~\ref{fig:KinematicDist2D}. Given these correlations, we can expect machine-learned observables to carry significant discriminating power in isolating (B)SM four top production.

%%%%%%%%%%%%%%%%%%%%%%%%%%%%%%%%%%
\subsection{Graph Neural Network Architecture and Data Representation}
%%%%%%%%%%%%%%%%%%%%%%%%%%%%%%%%%%
Given the complex final state that can be expected from four top topologies, in addition to the process-specific correlations of final state objects, GNNs are ideal candidate architectures to discriminate the signal characteristics from the expected background. Applications of GNNs to particle physics data have a short
yet successful history. Their versatility in efficiently discriminating BSM data from the SM expectation beyond traditional observables in a robust way has been highlighted in series of papers, ranging from designing anomaly detection methods~\cite{Blance:2020ktp,Atkinson:2022uzb,Atkinson:2021nlt}, over jet tagging~\cite{Dreyer:2020brq,Bols:2020bkb}  to constraining EFT operator deformations in the BSM top sector fits~\cite{Atkinson:2021jnj}.

Taking inspiration from this evolving success story, we employ a GNN to discriminate the four top signal from the relevant SM background in this work. To implement the graph structure, we use the \textsc{Deep Graph Library}~\cite{2019arXiv190901315W} with the \textsc{PyTorch}~\cite{NEURIPS2019_9015} backend and choose an Edge Convolution (EdgeConv) network to classify signal and background (final state particle events are interfaced employing \textsc{PyLHE}~\cite{pylhe}). 

The GNN can be divided into (i)~message passing followed by (ii)~node readout. Edge convolution is known to be particularly suited for extracting internode information (edge) from the given low-level node features (i.e. particle-level properties, see below).
The message passing function for the edge convolution is defined as
%%%%%%%%%%%%%%%%%%%%%
\begin{equation}
	\label{eq:edge_conv}
	\vec{x}_{i}^{\,(l+1)} = \frac{1}{|\mathcal{N}(i)|} \sum_{j \in \mathcal{N}(i)} {\text{\sc{ReLU}}}\left(\Theta \cdot (\vec{x}_j^{\,(l)} - \vec{x}_i^{\,(l)}) + \Phi \cdot (\vec{x}_i^{\,(l)})\right)\,.
\end{equation}
%%%%%%%%%%%%%%%%%%%%%
Here, $\vec{x}_i^{\,(l)}$ represents the node features of node $i$ in the $l$-th message passing layer, with $l=0$ denoting the input node features of the graph. %\ps{Remove highlighted input?}. 
The neighbourhood set $\mathcal{N}(i)$ consists of all nodes in the graph connected to node $i$. The linear layers $\Theta$ and $\Phi$ take the input vectors $(\vec{x}_j^{\,(l)}-\vec{x}_i^{\,(l)})$ and $\vec{x}_i^{\,(l)}$, respectively, and map them onto alternate dimension spaces where one applies an activation function on their vector sum. The dimensionality of the ``hidden'' feature space is chosen such that performance is optimised, while avoiding overfitting.

%%%%%%%%%%%%%%%%%%%%%%%%%%%%%%%%%
\begin{figure}[!t]
\centering
\includegraphics[width=1\textwidth]{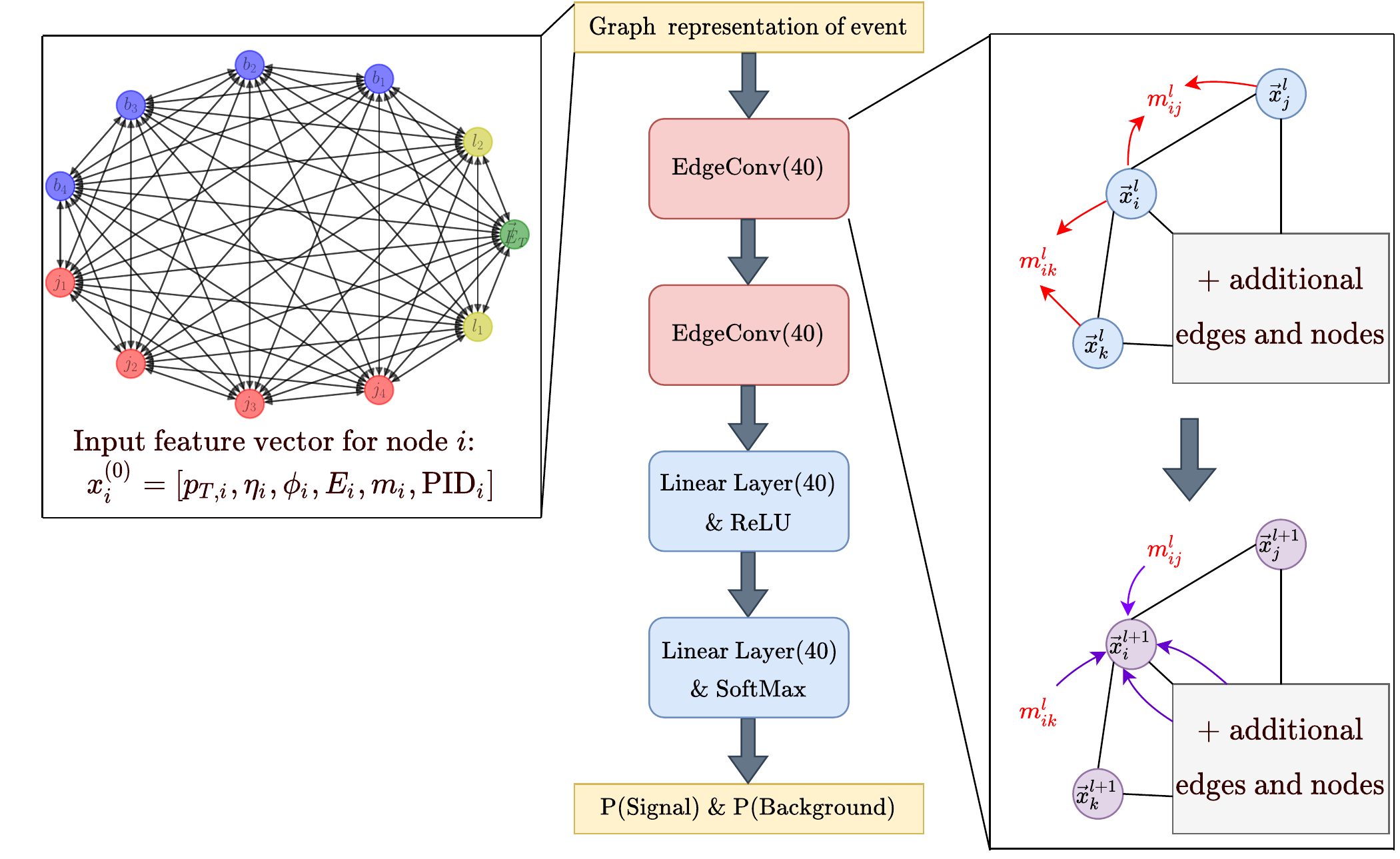}
	\caption{The architecture of the utilised GNN that maps the events embedded into graphs to the probability that they were sourced by signal or background processes. On the left, an example of an input graph is shown with different colours indicating the different final states that are included in our analyses. On the right the EdgeConv operation is indicated where during the first step (red arrows) the messages $m_{ij}^{(l)} = {\text{\sc{ReLU}}}\left(\Theta \cdot (\vec{x}_j^{\,(l)} - \vec{x}_i^{\,(l)}) + \Phi \cdot (\vec{x}_i^{\,(l)})\right)$ are calculated according to Eq.~\eqref{eq:edge_conv} and subsequently in the second step they are used to obtain the updated node features by taking the mean (purple arrows).  \label{fig:architecture}}
\end{figure}
%%%%%%%%%%%%%%%%%%%%%%%%%%%%%%%%%

For the four top vs. background discrimination task, we use a fully-connected bidirectional graph as input to the GNN. Each particle is represented by a node, which is associated with a node feature vector $[p_T$, $\eta$, $\phi$, $E$, $m$, PID$]$ (representing transverse momentum, pseudorapidity, azimuthal angle, energy, mass and particle identification number, respectively). 
For our analyses, after a fixed number of message passing steps Eq.~\eqref{eq:edge_conv}, we employ a mean graph readout operation to the node features of the final message passing layer 
to extract a vector capturing the graph properties of each event. The graph representation obtained this way is then fed into a linear layer with a \textsc{ReLU} activation function. The final linear layer maps the result to a two dimensional vector, normalised by the SoftMax function, which corresponds to the probability of an event arising from a signal or a background process. 

We use the \textsc{Adam} optimiser~\cite{2014arXiv1412.6980K} to minimise the cross-entropy loss function with an initial learning rate of 0.001. The network is trained separately with data corresponding to both final states. In each case, the data was split into a 75\%-25\% training-test ratio. The 75\% training data was further split into a 75\%-25\% training-validation ratio. The learning rate decays with a factor of 0.1 if the loss function has not decreased for three consecutive epochs. We train the models for 100 epochs in mini-batches of 100 graphs and an early-stopping condition on the loss for ten epochs. By using various combinations of EdgeConv and hidden linear layers, we identify two convolutional layers together with two linear layers and 40 nodes each as a suitable set of hyperparameters, as shown in Fig.~\ref{fig:architecture}. We verify that the validation and the training accuracies to check for overtraining (we find that deeper networks lead to overtraining by investigating accuracy and loss of the network).

Some comments regarding similar analysis strategies for the four top final state as given above are in order: a variety of ML algorithms have been explored in Ref.~\cite{Builtjes:2022usj} to assess their efficiency for the four top process, including the graph-based {\sc{ParticleNet}}~\cite{Qu:2019gqs} utilising the EdgeConv operation.\footnote{{\sc{ParticleNet}} uses the $k$-nearest-neighbours algorithm for the embedding of particles to graphs according to their separation $\Delta R = \sqrt{\Delta \eta^2 + \Delta \phi^2}$, while we rely on fully-connected graphs.} The authors of Ref.~\cite{Builtjes:2022usj} demonstrate that {\sc{ParticleNet}} does indeed provide superior performance in comparison with the alternative algorithms that are considered there. We do find a qualitatively similar behaviour.

%%%%%%%%%%%%%%%%%%%%%%%%%%%%%%%%%%
\section{Results}
\label{sec:results}
%%%%%%%%%%%%%%%%%%%%%%%%%%%%%%%%%%

%%%%%%%%%%%%%%%%%%%%%%%%%%%%%%%%%%
\subsection{Four tops at the LHC: Isolating the SM signal}
\label{sec:smsig}
The Standard Model four top final state has been explored by the ATLAS and CMS collaborations with considerable effort. 
ATLAS currently observes four tops with a sensitivity of $1.9\sigma$ over the background-only hypothesis
with a production cross section of 26 fb~\cite{ATLAS:2021kqb}. 
In comparison, CMS observes $2.6\sigma$ at a cross section of 12.6 fb~\cite{CMS:2019rvj}. 

\begin{figure}[!t]
\centering
\subfigure[~]{\includegraphics[width=0.48\textwidth]{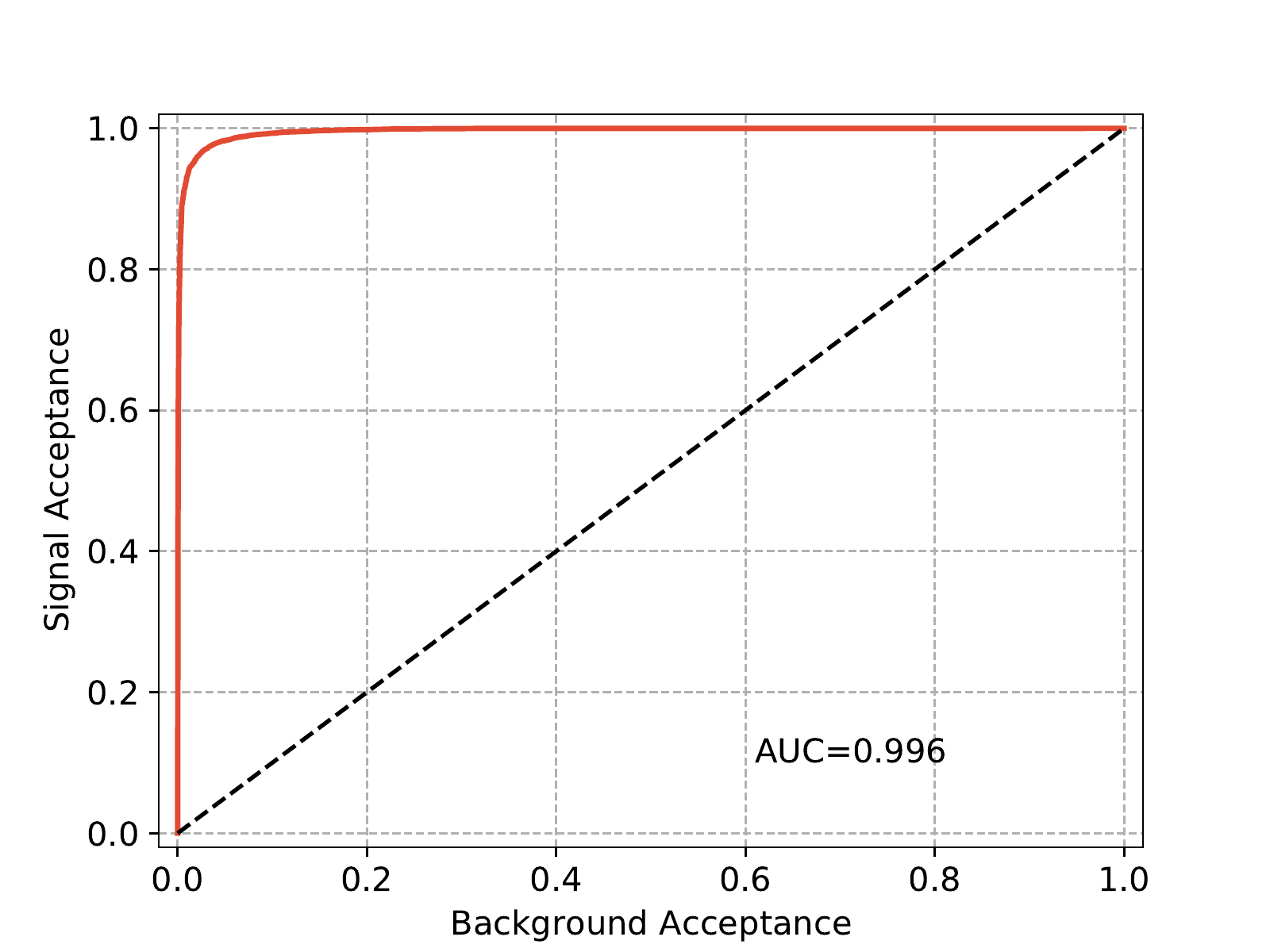}}\hfill
\subfigure[~]{\includegraphics[width=0.48\textwidth]{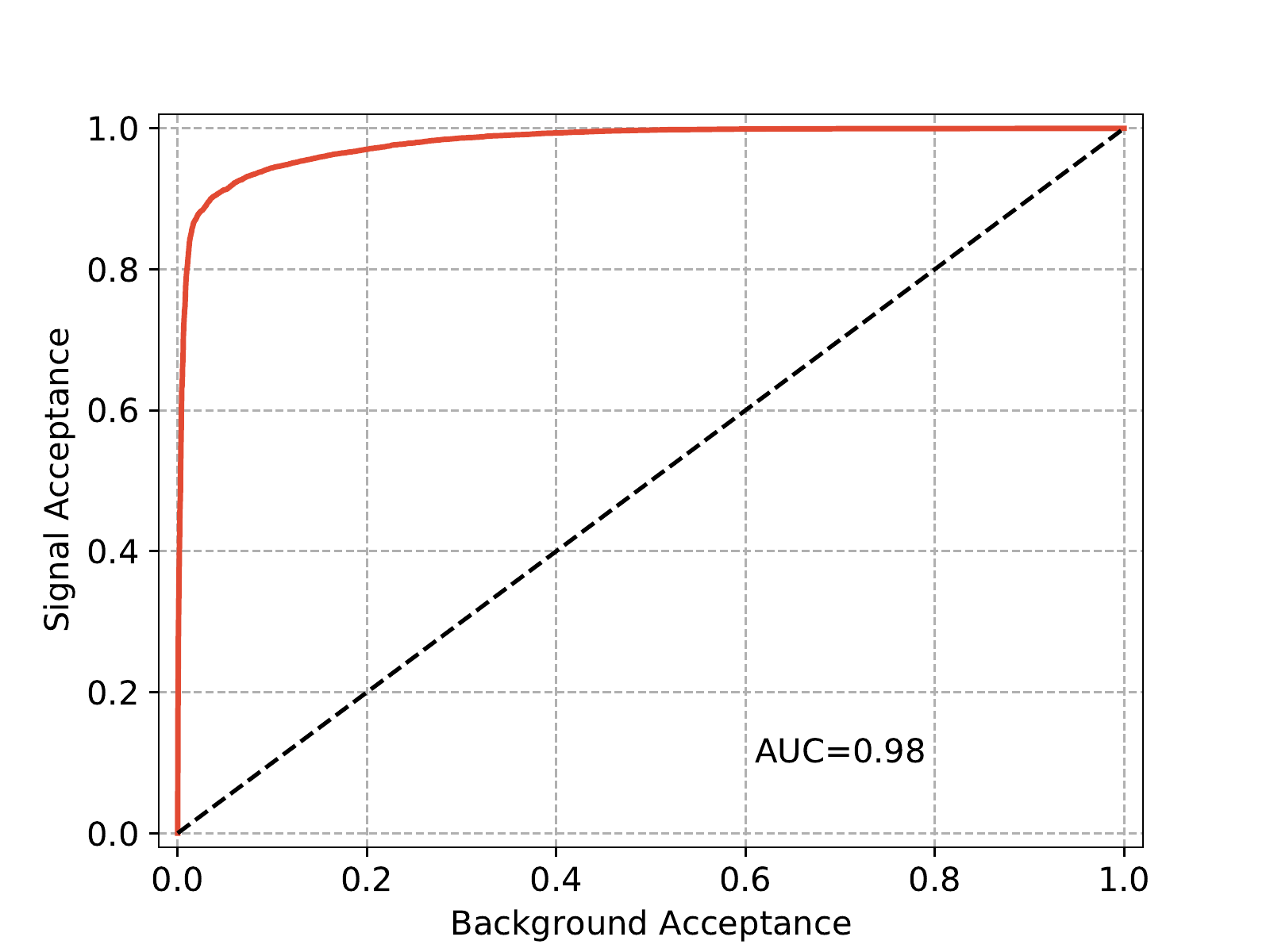}}
\caption{ROC curves for discriminating the SM four top quark signal in the (a) 2SSDL 
and (b)~3L 
final states.\label{fig:roc_sm}}
\end{figure}

To comment on the GNN performance for SM four top production, we first turn to the sensitivity estimates given by the backgrounds described in Sec.~\ref{sec:analysis} . We first train our network for the 2SSDL
samples for SM four top ($t\bar{t}t\bar{t}$) as the signal class and the corresponding backgrounds (listed in Tab.~\ref{tab:2ssdl_prod}) as the background class. 
A similar procedure is followed for the 3L 
final state. We demonstrate the network performance using the receiver operating characteristic (ROC) curves of the network which are shown in Fig.~\ref{fig:roc_sm} for the two different final states under consideration. The areas under the curves (AUCs) of 0.994 for 2SSDL 
and 0.968 for 3L, 
imply a very high classifier performance, as expected by the correlations of Figs.~\ref{fig:KinematicDist} and~\ref{fig:KinematicDist2D} (as well as the findings of Ref.~\cite{Builtjes:2022usj}). 
%%%%%%%%%%%%%%%%%%%%%%%%%%%%%%%%%%
%%%%%%%%%%%%%%%%%%%%%%%%%%%%%%%%%%

With the GNN setup optimised, we can cast the sensitivity encoded in the ROC curve into exclusion limits by considering the significance $\sigma={N_s}/{\sqrt{N_s+N_b}}$, where $N_s$ is the number of signal events and $N_b$ is the number of background events at a particular luminosity. The number of signal event $N_s$ is given by $\sigma^s\times \varepsilon_{s(GNN)}  \times L$ and number of background event is given is given by $N_B$ is given by $\sigma^b\times \varepsilon_{b(GNN)}  \times L$, where $\sigma^{s/b}$ is the cross section of signal and background after the baseline selection, $L$ is the integrated luminosity and $\varepsilon_{s/b(GNN)} $ are the efficiencies obtained at the best working point from the ROC curve such that $\sigma={N_s}/{\sqrt{N_s+N_b}}$ is maximised. 
We also explore the impact of systematic uncertainties by considering $20\%$ and $50\%$ overall systematic uncertainties in the analysis.

For the 2SSDL 
final state, the four top production cross section can be measured as $4.4~\text{fb}$ with a systematic uncertainty of $50\%$, and $4.1~\text{fb }$ with a systematic uncertainty of $20\%$, at an integrated luminosity of $139~\text{fb}^{-1}$ with a significance of $2.7\sigma$ over the background-only hypothesis using the GNN. Although we cannot claim comparability with the realistic experimental environments, these findings indicate that the GNN selection could indeed provide an avenue to further hone the sensitivity in ATLAS and CMS by integrating these techniques into their analysis flow. For the 3L 
final state, the four top production cross section can be measured as $18.6~\text{fb}$ with a systematic uncertainty of $50\%$, and $17.3~\text{fb}$ with a systematic uncertainty of $20\%$, at an integrated luminosity of $139~\text{fb}^{-1}$ with a significance of 2 standard deviations over the background-only hypothesis using GNN, implying a similar potential improvement as for 3L. The performance of the GNN in either channel reveals excellent discriminating power for the busy multi-top states at the LHC with prospects of improving cross section measurements significantly. 

%%%%%%%%%%%%%%%%%%%%%%%%%%%%%%%%%%
\subsection{Non-resonant new interactions: modified Higgs boson interactions}
\label{sec:nonres}
%%%%%%%%%%%%%%%%%%%%%%%%%%%%%%%%%%
Having demonstrated that the GNN indeed provides a suitable approach for extracting the four top SM signal, we can now turn to the relevance of the GNN for departures of the four top final states from the SM expectation. New physics modifications of four top final states from non-resonant interactions have been considered in a range of phenomenological studies~\cite{Banelli:2020iau,Aoude:2022deh}. To highlight the relevance of the above approach, we consider the
so-called $\hat{H}$ parameter as a motivated EFT-related example; Ref.~\cite{Englert:2019zmt} particularly emphasised the relevance of four top final states for associated searches (bounds have since been provided by the CMS experiment in~\cite{CMS:2019rvj}). Considering this particular non-resonant modification
is a well-motivated test bed to motivate a more comprehensive analysis of four top interactions from the perspective
of effective field theory~\cite{Atkinson:2021jnj,Aoude:2022deh}. 

The $\hat{H}$ parameter is the analogue of the $\hat W,\hat Y$ operators~\cite{Barbieri:2004qk} of the gauge sector and can be understood
as an oblique correction taking the form
\begin{equation}
\label{eq:hhat}
{\cal{L}}_{\hat H} = {\hat H\over m_H^2} |D_\mu D^\mu \Phi|^2\,,
\end{equation}
where $D_\mu$ is the covariant derivative action on the Higgs doublet $\Phi$.\footnote{It is worthwhile mentioning that the modification of the propagator
probes a genuine direction of Higgs Effective Field Theory (HEFT), that can be radiatively sourced in the so-called $\kappa$ framework~\cite{LHCHiggsCrossSectionWorkingGroup:2011wcg}, see e.g.~\cite{Herrero:2021iqt,Anisha:2022ctm}. The associated coupling modifications
can then be understood as a probe of the linearity of the Higgs boson's interactions in the SM and its dimension-6 extension~\cite{Grzadkowski:2010es}.} The presence of this interaction
modifies the propagation of the physical Higgs boson (suppressing the Higgs width which is not relevant as the Higgs boson is probed off-shell in four top production)
\begin{equation}
-i\Delta (p^2,m_H^2) = {1\over p^2 - m_H^2} - {\hat H\over m_H^2}
\end{equation}
with associated coupling modifications of the Higgs boson's couplings to massive vectors $V=W^\pm,Z$ and heavy fermions (here the top quark)
\begin{equation}
\label{eq:coupmod}
{g^{\hat H}_{VVH(p^2)} \over g^{\text{SM}}_{VVH}}= 1-\hat H\left(1-{p^2\over m_H}\right)\,,\quad
{g^{\hat H}_{t\bar tH} \over g^{\text{SM}}_{t\bar tH}}= 1-\hat H \,,
\end{equation}
for canonically normalised fields. The correlation of $HVV$ interactions and $H$ propagator resulting in a cancellation is 
a consequence of gauge symmetry~\cite{Englert:2019zmt}
\begin{equation}
 g^{\hat H}_{VVH} \Delta (p^2,m_H^2)=g^{\text{SM}}_{VVH} \Delta^{\text{SM}}_{\hat{H}=0} (p^2,m_H^2) + {\cal{O}}(\hat{H}^2)\,,
\end{equation}
which highlights multi-top final states as particularly suitable
process to put constraints on the interaction of Eq.~\eqref{eq:hhat} beyond Higgs coupling measurements. 

%%%%%%%%%%%%%%%%%%%%%%%%%%%%%%%%%%
\begin{figure}[!t]
\centering
\parbox{0.5\textwidth}{\includegraphics[width=0.48\textwidth]{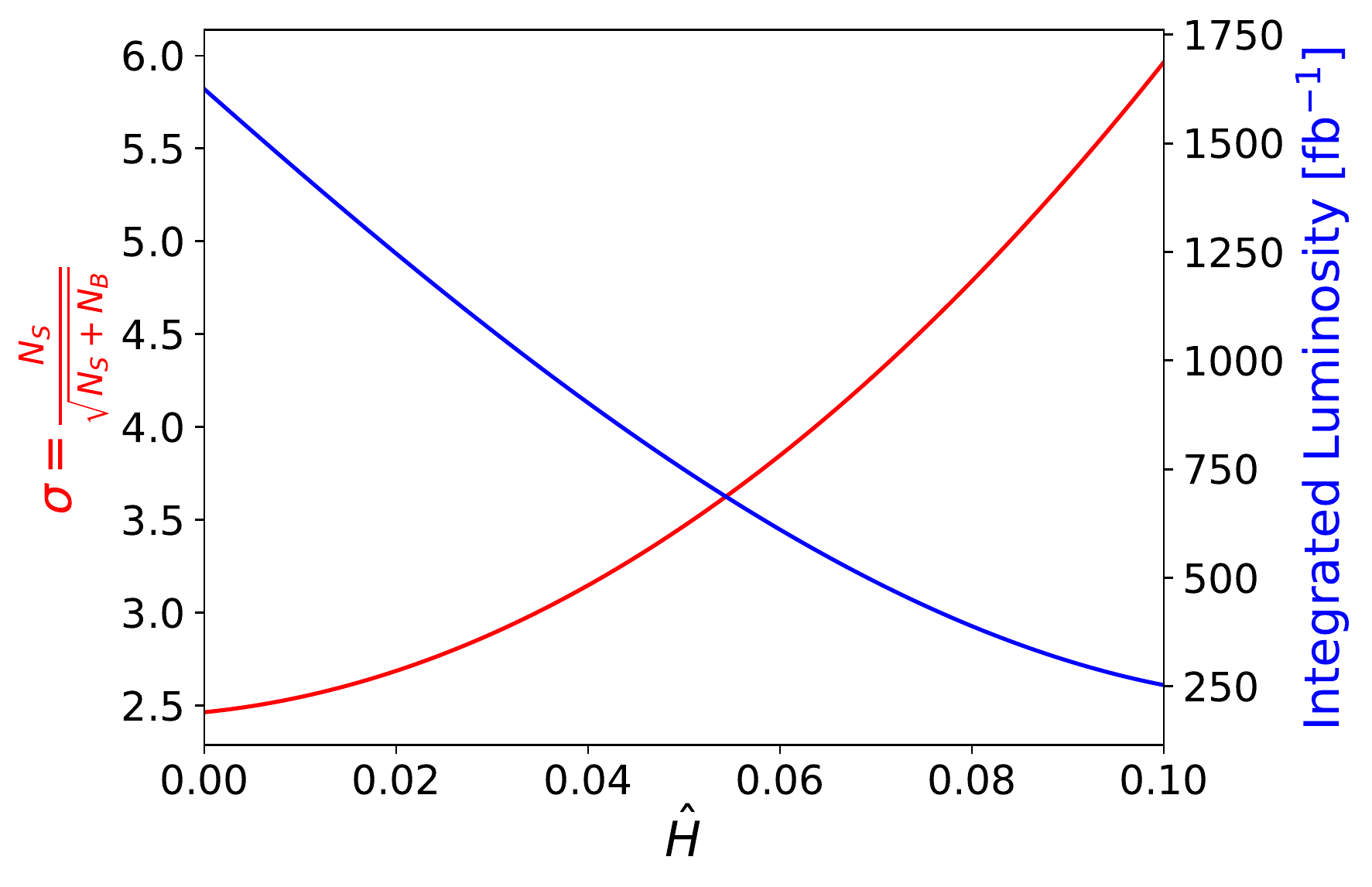}}
\hspace{0.4cm}
\parbox{0.42\textwidth}{
\vspace{1.72cm}\caption{Significance as a function of $\hat H$ at 13 TeV LHC with a $1000~\text{fb}^{-1}$
integrated luminosity. We also present the required luminosity as a $3\sigma$ confidence level for different $\hat H$ values. \label{fig:hhatlimit}}}
\end{figure}
%%%%%%%%%%%%%%%%%%%%%%%%%%%%%%%%%%

To include the effects of $\hat{H}$ in our study, we modify the {\sc{Helas}} routines~\cite{Murayama:1992gi,deAquino:2011ub} of our $pp\to t\bar t t \bar t$ implementation to reflect the modifications of propagators, vertices and their cancellation to linear order in the amplitude. As a benchmark for non-resonant new interactions, we change $\hat H$ = 0.1 as a new physics modification (this value is motivated from the HL-LHC measurement sensitivity provided in Ref.~\cite{Englert:2019zmt}). We select a SM-rich sample by further applying a cut on $H_T$ greater than 900 GeV to enable a coupling measurement that is more geared to SM modifications, see Fig.~\ref{fig:KinematicDist}. We then train a GNN network to discriminate between SM interaction with $\hat H$ = 0 from the non-resonant new interaction with $\hat H$ = 0.1. The kinematical differences are not dramatic (which is highlighted by the $\sim 10\%$ sensitivity in the first place), and therefore further motivate the inclusion of as much correlated information as possible achievable through GNN applications. 

We find an AUC of $60\%$, which is testimony to the difficulty of extracting electroweak properties from QCD-busy final states.
The observed sensitivity can be used to set limits for in the parameter range $\hat H \in [0.0, 0.10]$ shown in Fig.~\ref{fig:hhatlimit}. The blue curve shows the required luminosity to obtain the $3\sigma$ bound on that particular $\hat H$ value, whereas the red curve shows the significance obtained at a fixed luminosity $L=1000~\text{fb}^{-1}$. Our GNN analysis again gives rise to an expected sensitivity improvement beyond the estimates of cut-and-count analyses~\cite{Englert:2019zmt,CMS:2019rvj}. While the realistic analysis in the LHC environment is likely more limited in sensitivity, the addition of the GNN techniques described above could significantly improve the sensitivity to $\hat H$ considerably, which also indicates an improved sensitivity to generic EFT deformations.
%%%%%%%%%%%%%%%%%%%%%%%%%%%%%%%%%%
\begin{figure}[!t]
\centering
\includegraphics[width=0.48\textwidth]{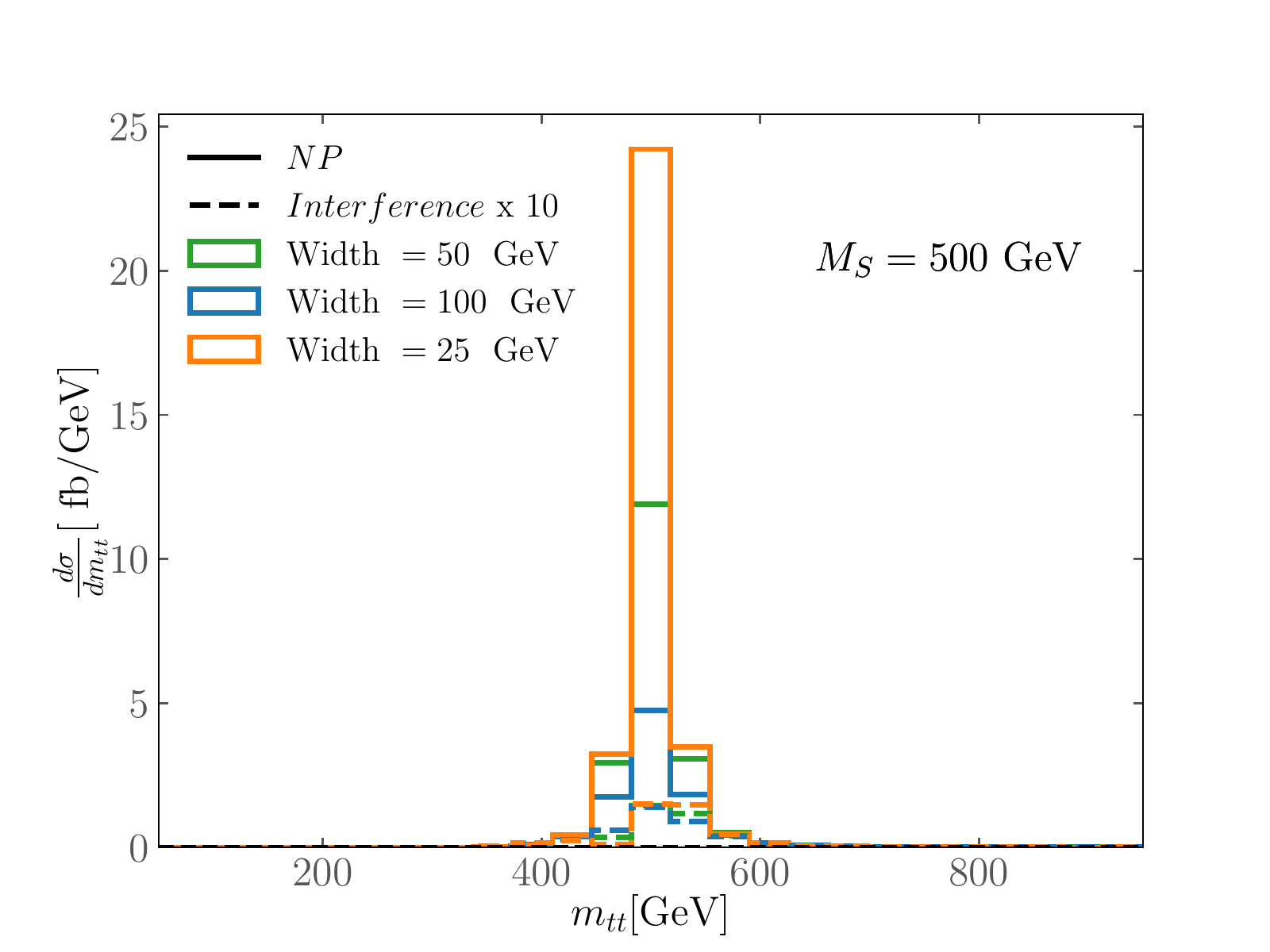}
\includegraphics[width=0.48\textwidth]{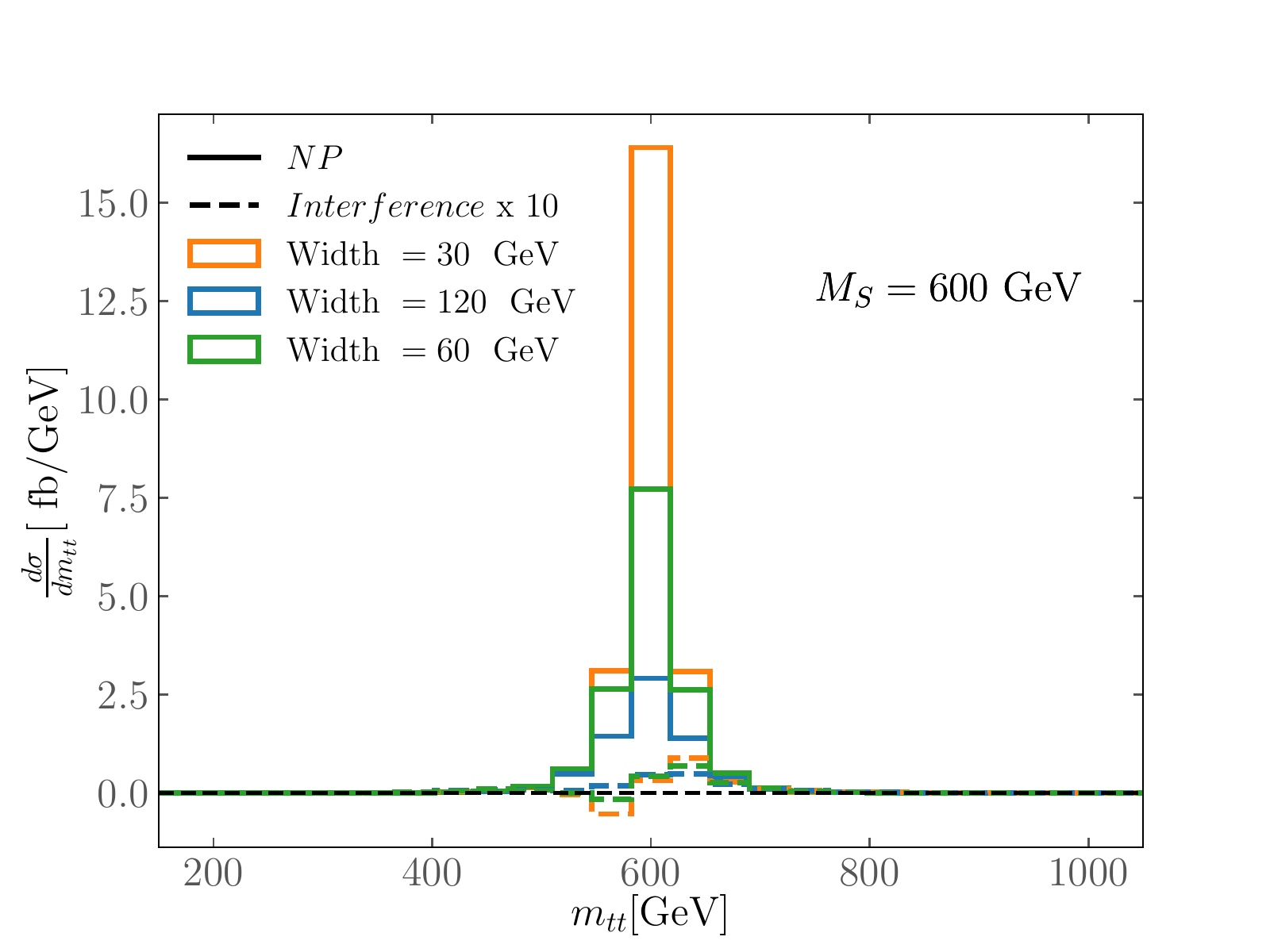}\\
\includegraphics[width=0.48\textwidth]{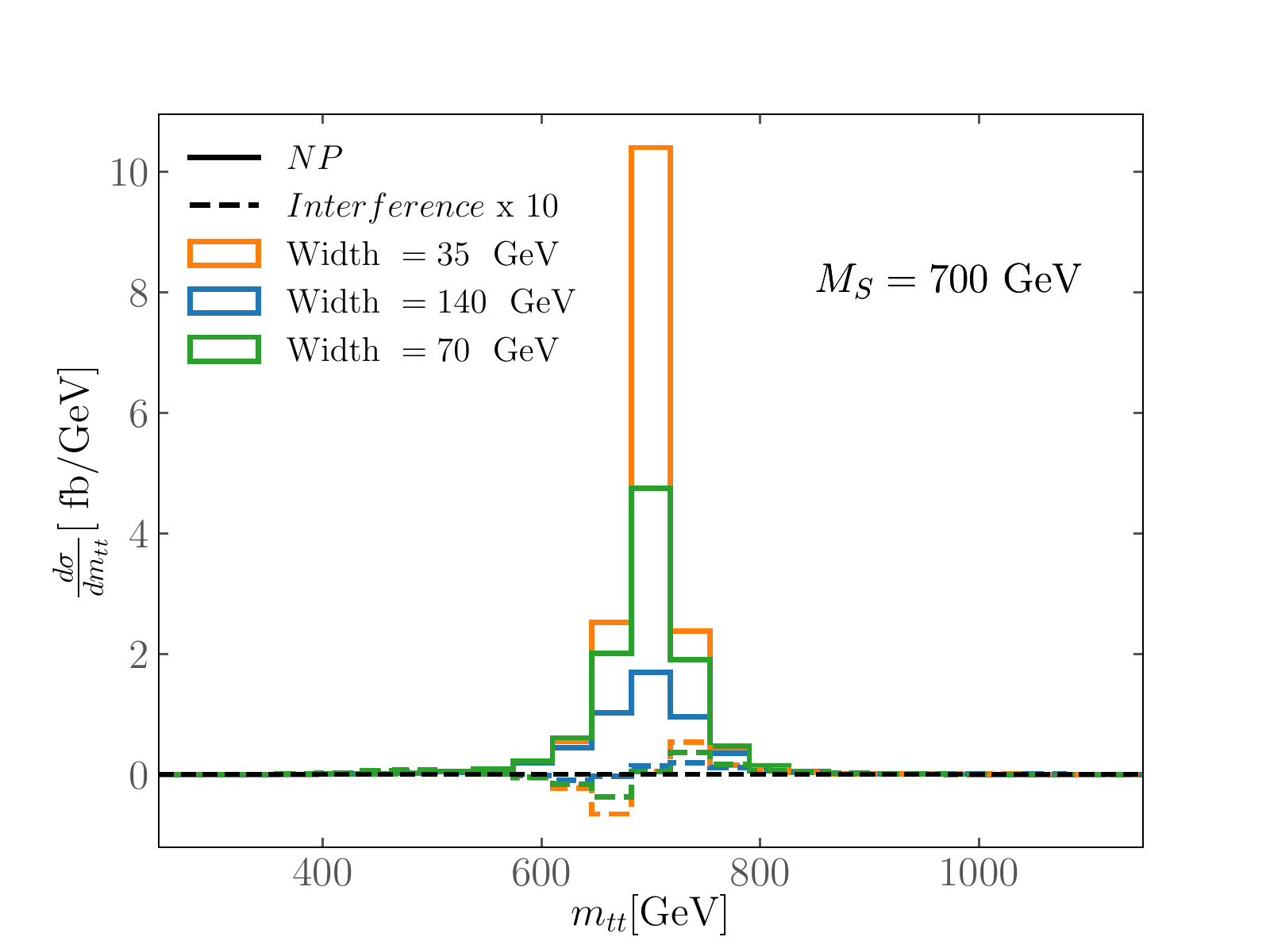}
\includegraphics[width=0.48\textwidth]{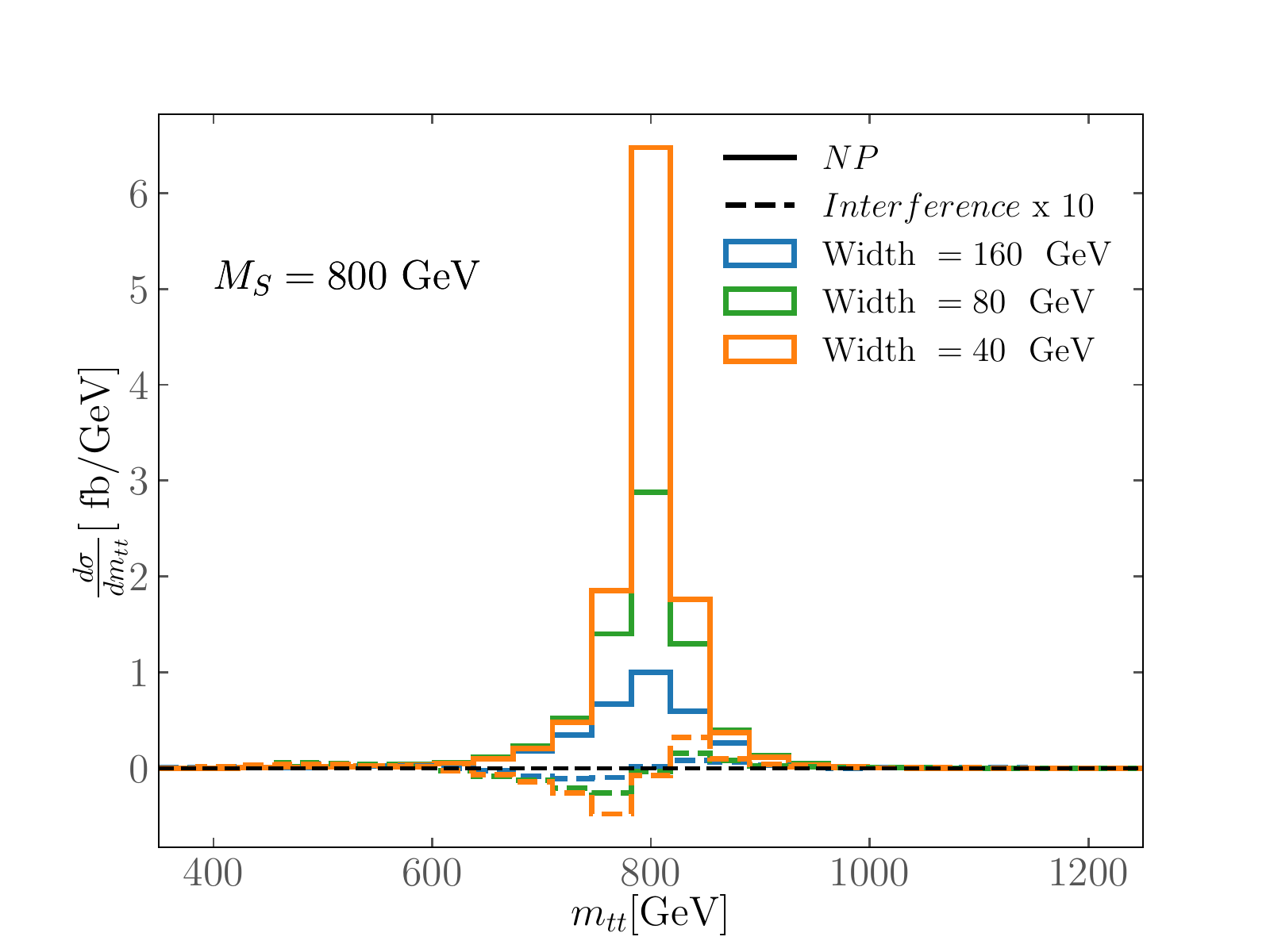}\\
\includegraphics[width=0.48\textwidth]{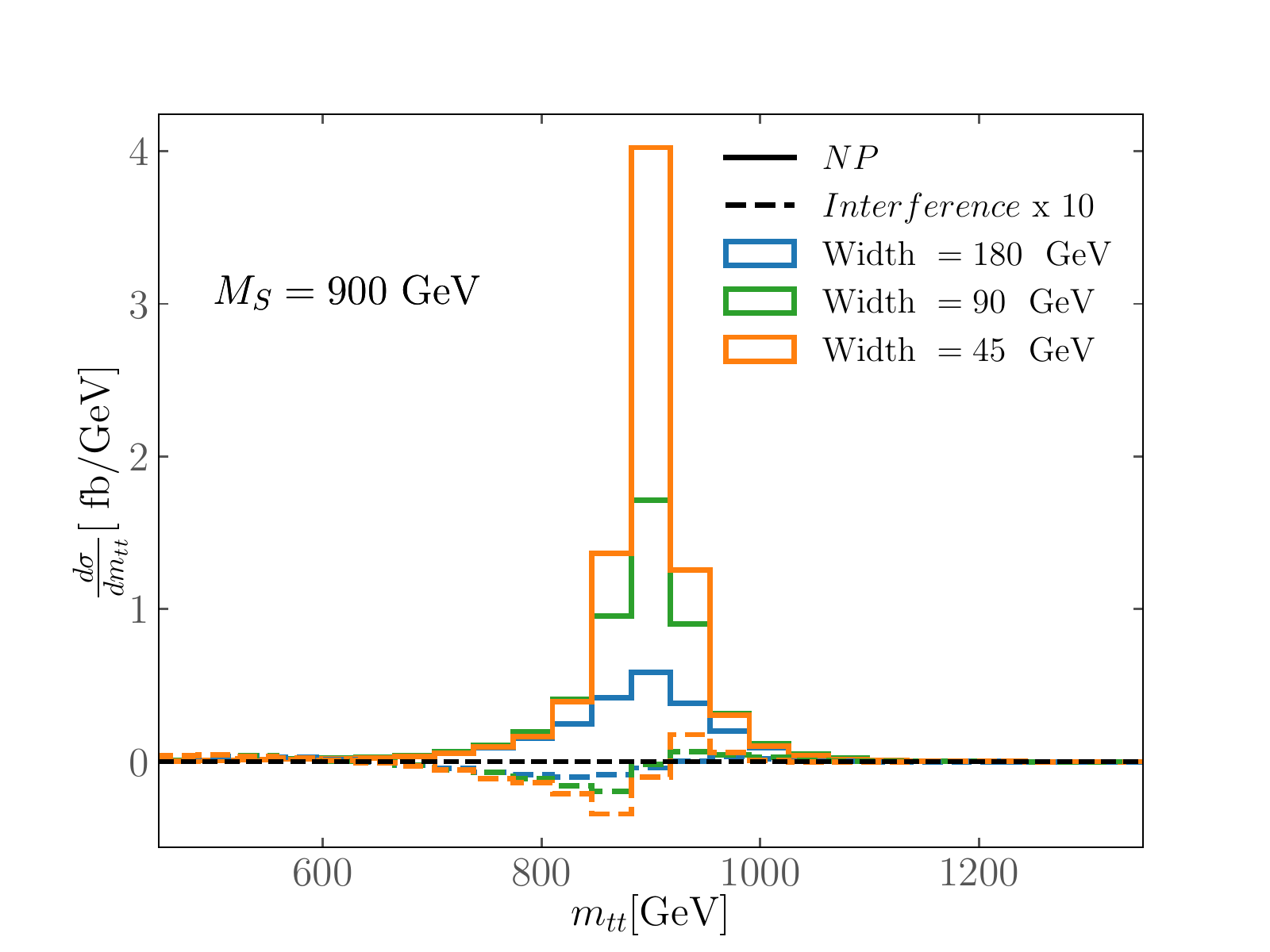}
\includegraphics[width=0.48\textwidth]{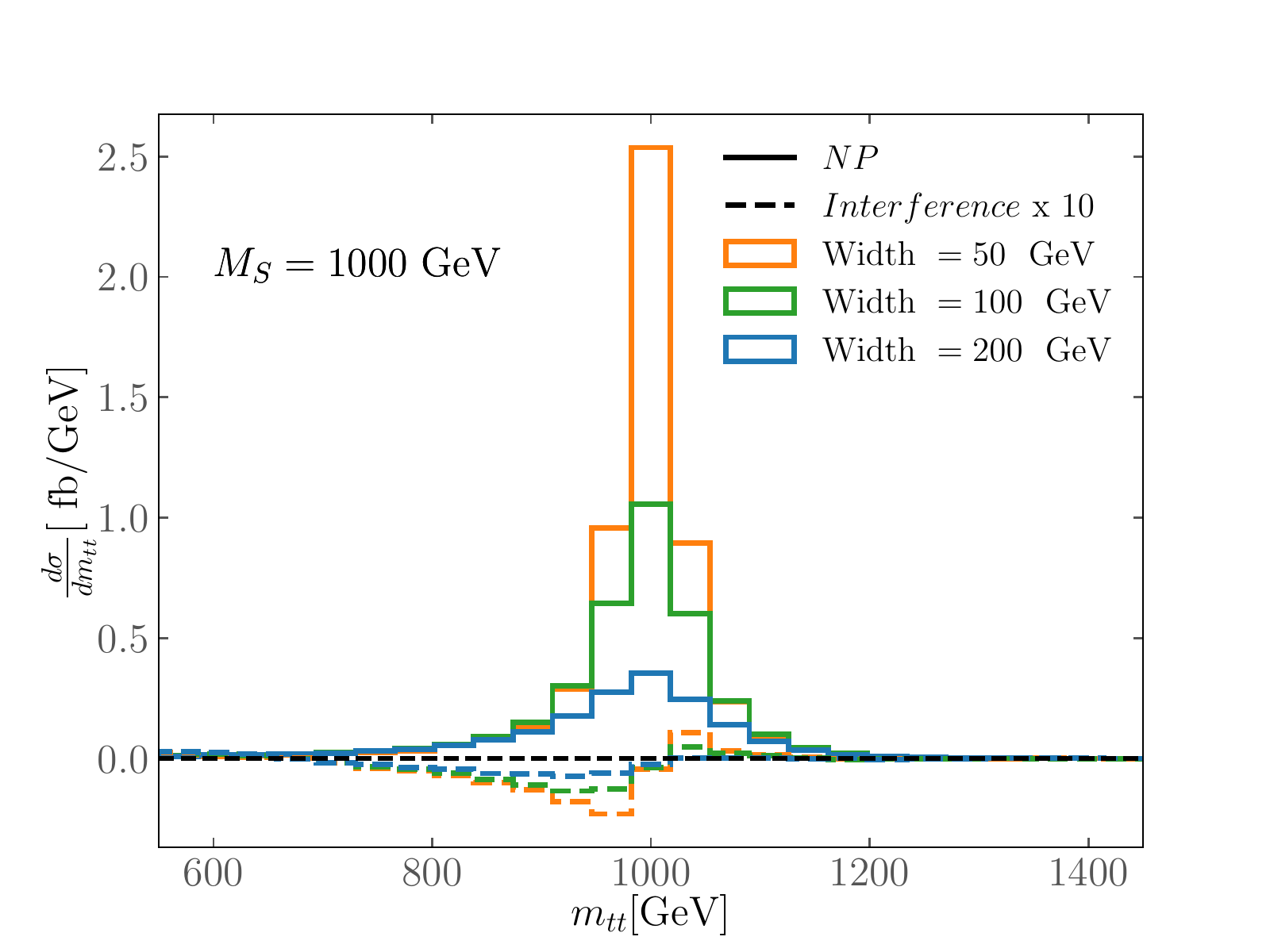}
\caption{The invariant mass distribution of $m_{tt}$ system closest to the mass of the new scalar $S$ as detailed in the text.  \label{fig:interference}}
\end{figure}
%%%%%%%%%%%%%%%%%%%%%%%%%%%%%%%%%%

%%%%%%%%%%%%%%%%%%%%%%%%%%%%%%%%%%
\subsection{Resonant new interactions: interference effects and 2HDM reach}
\label{sec:res}
%%%%%%%%%%%%%%%%%%%%%%%%%%%%%%%%%%
Four top final states in the context of resonant extensions of the SM have been studied in a range of analyses~\cite{ATLAS:2022ohr, Blekman:2022jag,Alvarez:2019uxp,Alvarez:2016nrz,Kanemura:2015nza}, typically relying on traditional collider observables alongside cut-and-count strategies. Whilst new resonant structures lend themselves to
such approaches, we can also expect a significant sensitivity enhancement when turning to ML approaches (see also~\cite{Butter:2017cot,Collins:2019jip,Atkinson:2020uos,Freitas:2020ttd,Hollingsworth:2020kjg,Kitouni:2020xgb}). 

Before we comment on the GNN performance, however, we will perform a qualitative analysis of the interference effects
that are known to be large in direct gluon fusion production $gg\to t\bar t$ of extra scalar resonances~\cite{Gaemers:1984sj,Dicus:1994bm}.
Such interference effects can severely limit the sensitivity estimates and are not straightforward to include in ML-based selections due to negative weights.
To this end we introduce a simplified scalar resonance $S$ with Higgs-like couplings to the top quark\footnote{We model the phenomenological Lagrangian with {\sc{FeynRules}}~\cite{Christensen:2008py,Alloul:2013bka} and export the relevant Feynman Rules in the {\sc{Ufo}}~\cite{Degrande:2011ua} format. We use the same toolchain as before with the addition of {\sc{MadSpin}}~\cite{Frixione:2007zp,Artoisenet:2012st}.}
\begin{equation}
\label{eq:simp}
{\cal{L}}_{\text{simp}} = {1\over 2} (\partial S)^2 - {M_S^2\over 2} S^2  - {m_t\over v} \left[ \xi_S \,\bar t_L t_R S + {\text{h.c.}}  \right]\,,
\end{equation}
keeping the width of $S$ as a free parameter in a scan to qualitatively assess $S$ resonance-distortion ($\xi_S$ plays the role of the Higgs coupling modifier for $S=H$). 
To study the interference effects in the four top final states, we compare the resonance structure stemming for the new-physics-only (resonance) interactions  
\begin{equation}
\hbox{d}\sigma^{\text{new}} \sim |{\cal{M}}_{\text{res}}|^2~\text{dLIPS} 
\end{equation}
with interference-only contributions
\begin{equation}
\hbox{d}\sigma^{\text{inf}} \sim  2\,\text{Re}\left( {\cal{M}}_{\text{bkg}} \,{\cal{M}}_{\text{res}}^\ast \right)\text{dLIPS}\,,
\end{equation}
where `bkg' refers to any non-resonant amplitude contribution (e.g. the continuum QCD background).
A good qualitative understanding of the signal distribution distortion can be obtained by isolating the peaked $m_{t\bar t}$ distribution: we construct all possible combinations of $m_{t\bar{t}}$ and choose the invariant mass which is closest to the candidate mass $M_{S}$. The resulting distributions are shown in Fig.~\ref{fig:interference} for a range of mass $M_{S}=[500, 600, 700, 800, 1000]~\text{GeV}$ and width choices $\Gamma_S/M_S=5\%,10\%,20\%$ for $\xi_S=1$.

From the plots shown in Fig.~\ref{fig:interference}, we can infer that the interference effects that distort the mass peak are comparably small. Closer to the decoupling limit $\xi_S\to 0$ we will encounter more sensitivity-limiting distortion, however this will happen at quickly vanishing BSM resonance cross sections so that a (ML-assisted) bump hunt will not provide any sensitivity when considering the backgrounds. 
In the case of imaginary phases of $\xi_S$, i.e. CP-odd coupling structures, the interference effects become decreasingly relevant as we move towards CP-odd states like those predicted in the two Higgs doublet model (2HDM). This is due to the dominant QCD background being characterised by CP-even interactions and any interference between CP-odd and CP-even amplitudes cancels in CP-even mass distributions unless resolved specifically with CP-sensitive observables (see also~\cite{Atkinson:2020uos}). 

\begin{figure}[!ht]
\centering
\subfigure[]{\includegraphics[width=0.48\textwidth]{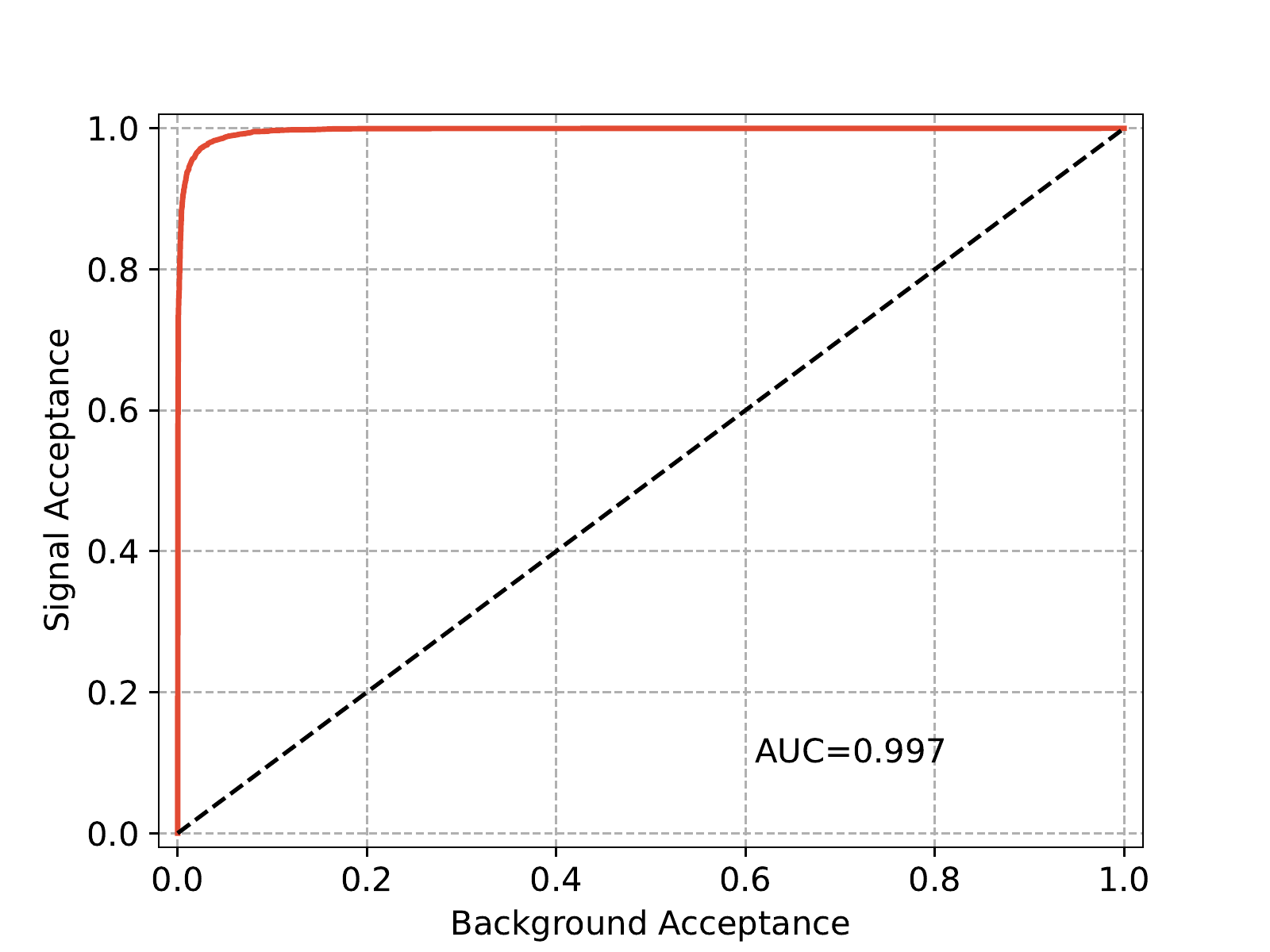}}\hfill
\subfigure[]{\includegraphics[width=0.48\textwidth]{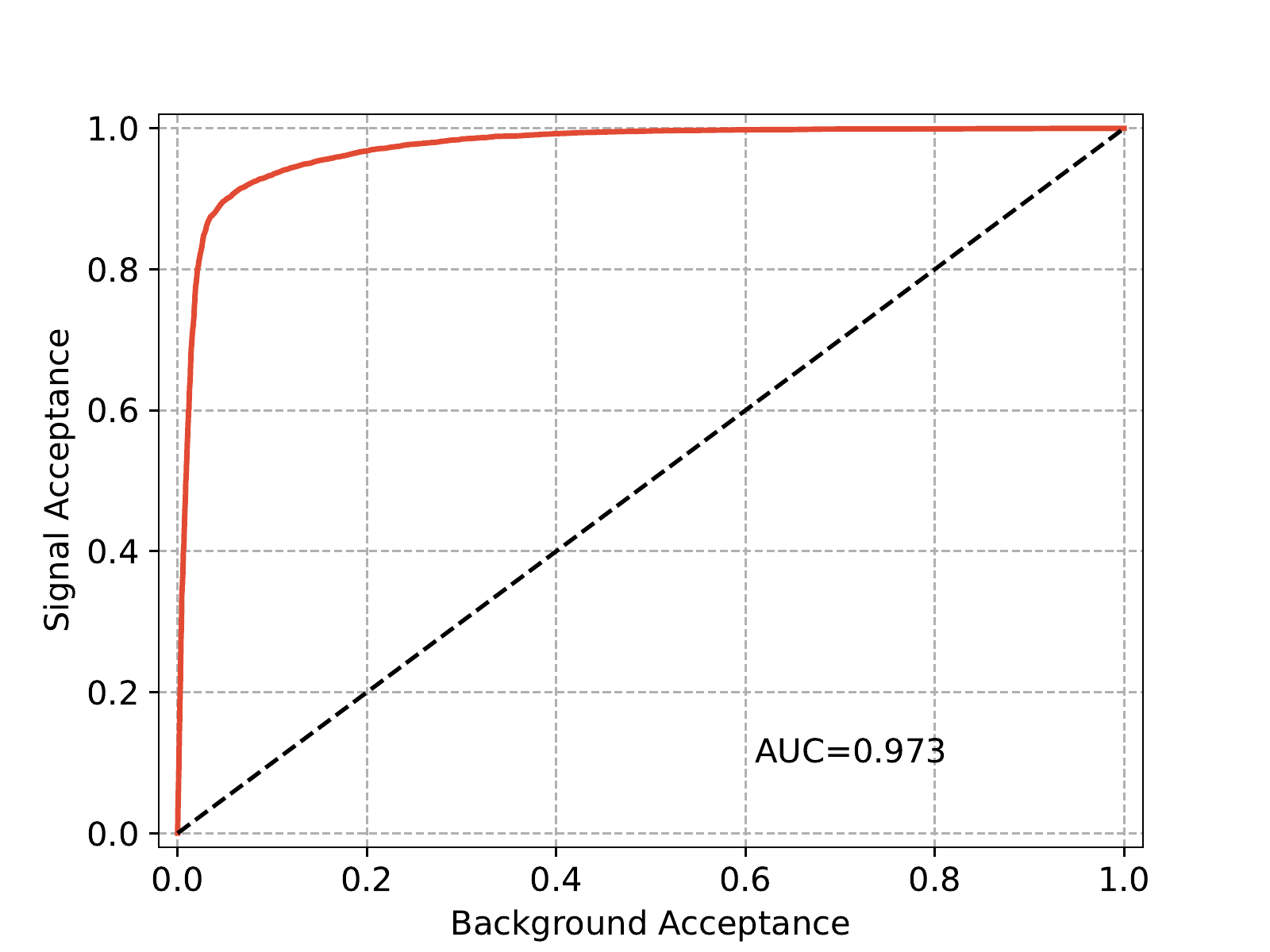}}
\caption{\label{fig:roc_bsm} ROC curves for training a four top quark signal for a BSM CP-even scalar with $M_S=600~\text{GeV}$ for (a)~2SSDL and (b)~3L final states.}
\end{figure}

\begin{figure}[!b]
\centering
\parbox{0.5\textwidth}{\includegraphics[width=0.48\textwidth]{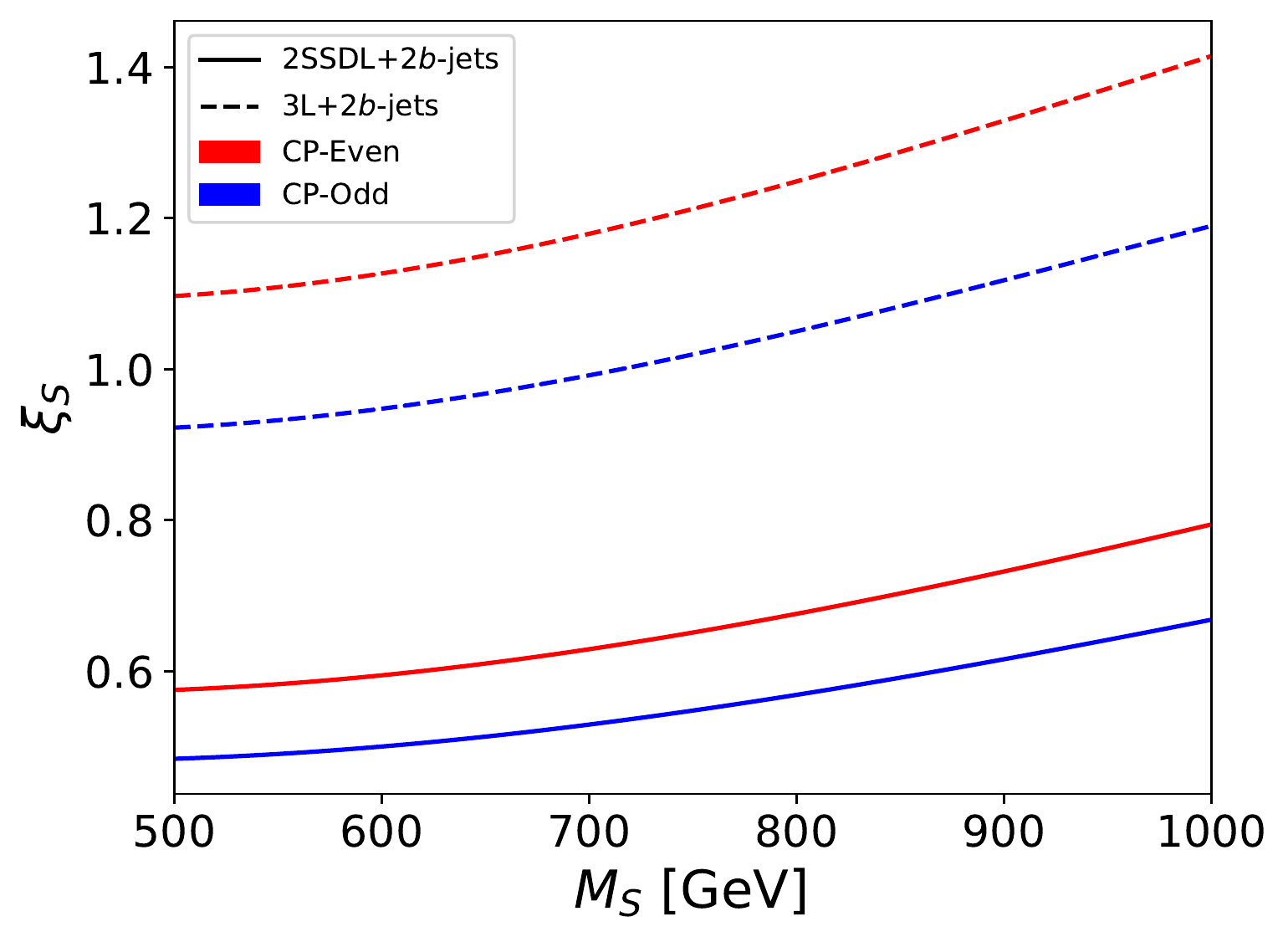}}
\hspace{0.4cm}
\parbox{0.42\textwidth}{
\vspace{2cm}
\caption{$2\sigma$ confidence level limits on scalar mass with respect to CP-even and CP-odd couplings for $3000~\text{fb}^{-1}$ for 13~TeV LHC collisions. \label{fig:limityukawa}}}
\end{figure}

Having identified interference effects as largely irrelevant, we return to the GNN analysis of the BSM resonance structures. We train our network for the same two final states, this time with the added new interactions (i.e., the newly added CP-even scalar with $\xi_S=1$) for the different benchmark points listed in Tab.~\ref{tab:limitsbsm}. The ROC curves for one such benchmark point ($M_S=600~\text{GeV}$) for both the final states and their corresponding AUCs are shown in Fig.~\ref{fig:roc_bsm}. The corresponding significance for each benchmark  point for the different decay channels is also listed in Tab.~\ref{tab:limitsbsm} for $\xi_S=1$ and $\Gamma_S/M_S=0.1$. We interpret these limits as bounds on the coupling modifier $\xi_S$ as shown in Fig.~\ref{fig:limityukawa}. 

%%%%%%%%%%%%%%%%%%%%%%%%%%%%%%%%%%
%%%%%%%%%%%%%%%%%%%%%%%%%%%%%%%%%%
%%%%%%%%%%%%%%%%%%%%%%%%%%%%%%%%%%
\begin{table}[!ht]
\centering
\begin{tabular}{|c|c|c|}
\hline\hline
$M_S$ [GeV]           & significance for 2SSDL  & significance for 3L   \\ \hline\hline
500&   $10.6\sigma  $ &$ 1.4 \sigma $\\ 
600&  $9.9\sigma  $ &  $1.3\sigma $\\ 
700& $8.4\sigma    $ & $1.1 \sigma$\\
800&  $6.9 \sigma   $&  $0.9\sigma$\\
900&   $5.5  \sigma $ &$0.7\sigma$ \\
1000&  $ 4.3 \sigma$  & $0.5\sigma$\\
\hline\hline
\end{tabular}
\caption{\label{tab:limitsbsm} Significances for different masses $M_S$ for a coupling choice $\xi_S=1$ and  $\Gamma_S/M_S=0.1$, Eq.~\eqref{eq:simp}. The luminosity is taken to be $3000~\text{fb}^{-1}$ for 13 TeV LHC collisions. }
\end{table}
%%%%%%%%%%%%%%%%%%%%%%%%%%%%%%%%%%
%%%%%%%%%%%%%%%%%%%%%%%%%%%%%%%%%%
%%%%%%%%%%%%%%%%%%%%%%%%%%%%%%%%%%

%%%%%%%%%%%%%%%%%%%%%%%%%%%%%%%%%%
\begin{figure}[!t]
\centering
\parbox{0.5\textwidth}{
\includegraphics[width=0.5\textwidth]{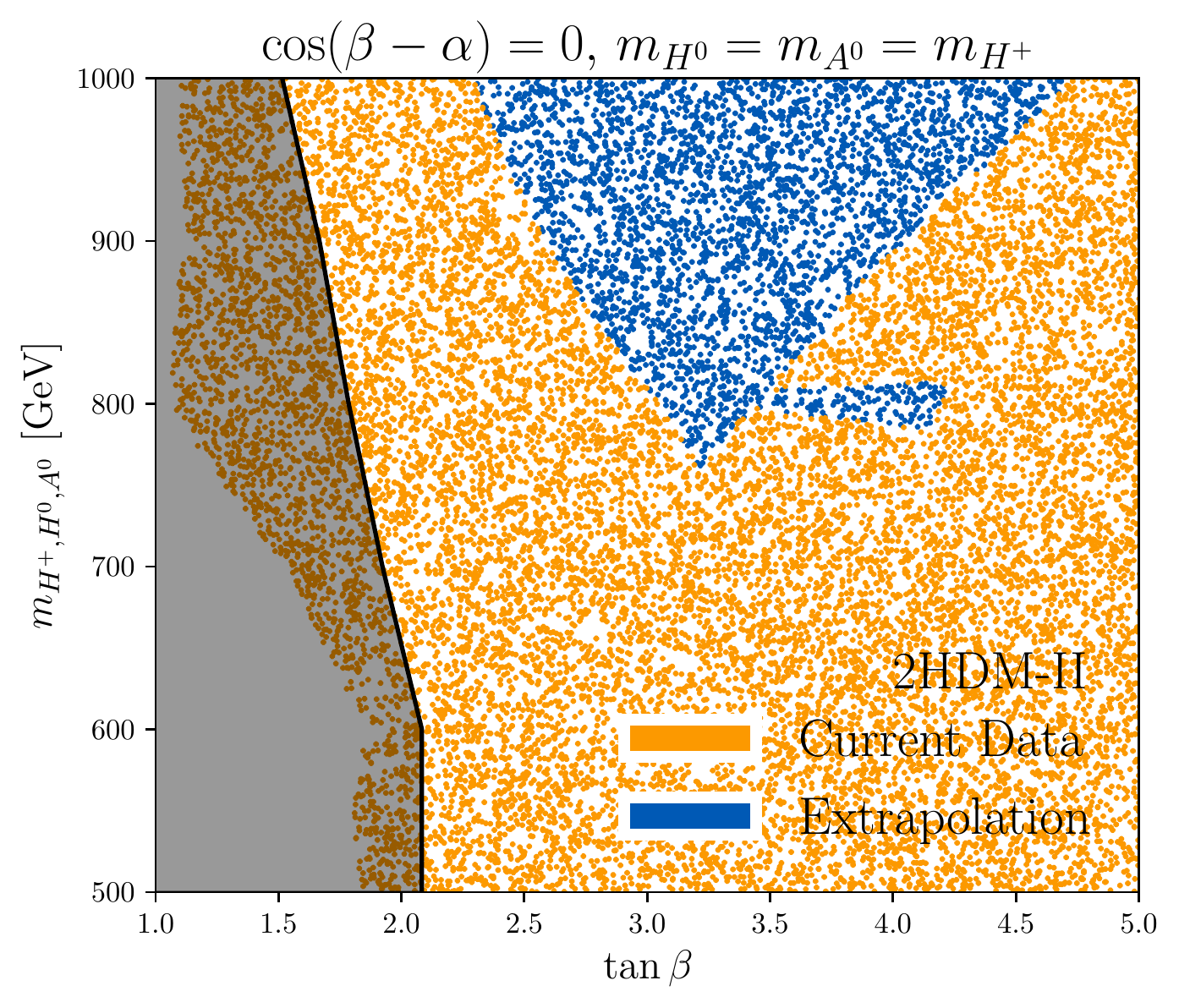}
}
\hspace{0.4cm}
\parbox{0.42\textwidth}{
\vspace{0.2cm}\caption{Scan of the parameter space of the type II 2HDM in the face of current collider data, and LHC search data extrapolated to $3000~\text{fb}^{-1}$ at 13~TeV, with a $2\sigma$ confidence level exclusion contour from 2SSDL final states. Points in blue are allowed by current and expected future datasets, orange points by current data only. The shaded area gives the sensitivity of the 2SSDL resonance search detailed in the text at $3000~\text{fb}^{-1}$. Interference effects from $gg\to t\bar t$ sensitivity estimates are not included. \label{fig:2hdm}}}
\end{figure}
%%%%%%%%%%%%%%%%%%%%%%%%%%%%%%%%%%

Although we can indeed put constraints $\xi_S<1$ (predominantly through the 2SSDL channel), the resulting bounds are relatively weak. For instance in singlet mixing scenarios, the parameter range that this analysis is sensitive to is already highly constrained from Higgs signal strength measurements. Turning to phenomenologically richer scenarios we further examine the 2HDM of type II, which features new neutral CP odd and CP even states. In this model, the Higgs sector consists of two distinct complex doublets, both with a non-zero vacuum expectation value (vev) (reviews can be found in Refs.~\cite{Branco:2011iw,Gunion:1989we}). One of these doublets couples to the up-type quarks and the other to down-type quarks and leptons in the type II model, thereby leading to fermion mass terms. After electroweak symmetry breaking, one is left with 5 Higgs bosons; 4 of which are CP even; $h$, $H$ and $H^\pm$, and a CP odd particle, $A$. The lighter of the two neutral CP even states is identified as the observed Higgs boson. The phenomenology of this model then depends on the masses of the new Higgs bosons, the ratio of the two vevs, defined as $\tan\beta = v_2/v_1$ and $\cos(\beta - \alpha)$, where $\alpha$ is the mixing angle between the two neutral CP even states. The simplified Lagrangian of Eq.~\eqref{eq:simp} can be mapped onto the 2HDM type II Yukawa sector
\begin{equation}
    {\cal L}_{\text{2HDM}}  \supset  
    - \frac{m_t}{v} 
    \left(\xi_h \, \bar{t} t h + \xi_H \, \bar{t} t H - i \xi_A  \, \bar{t} \gamma^5 t A \right)\,,
\end{equation}
where we have suppress charge Higgs contributions as well as non-top interactions. For the 2HDM type the coupling modifiers related to the top quark are
\begin{subequations}
\begin{align}
\xi_h &= \sin(\beta-\alpha) + \cos(\beta-\alpha)\cot\beta\,,\\
\xi_H &= \cos(\beta-\alpha) - \sin(\beta-\alpha)\cot\beta\,,\\
\xi_A &= \cot\beta\,.
\end{align}
\end{subequations}
The 2HDM is a particularly well motivated BSM model, as it is, in principle, capable of resolving a number of tensions between current experimental values and SM predictions, as well as opening avenues to satisfy the Sakharov criteria for required baryogenesis~\cite{Sakharov:1967dj}. As such, this model is well studied in the literature~\cite{Basler:2016obg,Kling:2016opi,Arnan:2017lxi,Crivellin:2019dun,Han:2020zqg,Atkinson:2021eox,Atkinson:2022pcn,Biekotter:2022kgf,Dawson:2022cmu}. Here we use this model to explore the validity of the benchmark points examined above. In order to satisfy the theoretical considerations the masses of the new particles are set to be entirely degenerate. Similarly, to fall in line with the signal strength data of the SM Higgs we set $\cos(\beta - \alpha) = 0$; the alignment limit which recovers exactly the SM phenomenology of $h$. 

To examine the mass range of the above benchmark points in the face of the collider searches for additional Higgs bosons we make use of the \texttt{2HDecay}~\cite{Krause:2018wmo, Djouadi:1997yw, Djouadi:2018xqq, Krause:2016xku, Denner:2018opp, Hahn:1998yk} and \texttt{HiggsBounds}~\cite{Bechtle:2008jh, Bechtle:2011sb, Bechtle:2012lvg, Bechtle:2013wla, Bechtle:2015pma, Bechtle:2020pkv, Bahl:2021yhk} packages. Additionally, we extrapolate the LHC search data to $3000~\text{fb}^{-1}$ at 13~TeV. Performing a scan over 20000 random points, the results of this analysis are shown in Fig.~\ref{fig:2hdm}, with points in blue allowed by current and extrapolated data, while orange points are allowed only by the current data. The improvement in sensitivity is lead by searches for the decays $H^+ \to t\bar{b}$ and $H^0 \to \tau^+\tau^-$ \cite{CMS:2020imj, ATLAS:2020zms}. This shows that a large swathe of the parameter space that we take for benchmark points in the above analysis is possible within the framework of the 2HDM of type II. 

Given that four top final states are far less abundant than other processes, they can be expected to exhibit a reduced sensitivity to new states (this is already visible from Fig.~\ref{fig:limityukawa}). Nonetheless, repeating the above analysis also including the CP odd state, we find that complementary sensitivity can be achieved for the 2HDM scenario with the GNN-assisted resonance search detailed above: By overlaying the constraints found from 2SSDL final states (shaded area) we see that there is a non-trivial parameter range of the 2HDM type II's alignment limit that can be accessed four top final states at $2\sigma$ confidence and $3000~\text{fb}^{-1}$. This example contains only the more sensitive channel of the ones discussed in this work and additional final states are likely to further improve this result.

%%%%%%%%%%%%%%%%%%%%%%%%%%%%%%%%%%
\section{Summary and Conclusions}
\label{sec:conc}
%%%%%%%%%%%%%%%%%%%%%%%%%%%%%%%%%%
The growing sensitivity to four top final states at the Large Hadron Collider~\cite{CMS:2019rvj,ATLAS:2020hpj,ATLAS:2021kqb,CMS:2022uga,CMS:2019jsc,ATLAS:2018kxv} bears great potential for the discovery of new physics beyond the Standard Model. While the complexity of the final states requires a dedicated search strategy compared to lower multiplicity processes, the plethora of kinematic information that can be accessed in these processes to highlight departures from the SM can be formidably exploited by means of adapted search strategies, e.g. building on Machine Learning approaches. These do not only provide support in establishing sensitivity to SM four top production as shown in Sec.~\ref{sec:smsig} (see also~\cite{Builtjes:2022usj}), but they become particularly relevant when four top final states will be scrutinised from a BSM perspective in the future. We have demonstrated that significant sensitivity to well-motivated representative non-resonant physics can be achieved (Sec.~\ref{sec:nonres}), but also that sensitivity to resonances is gained by turning to Graph Neural Networks (Sec.~\ref{sec:res}). The GNN exploits the hierarchical graph structure of particle physics events that is particularly accentuated when dealing with new resonanes. In this case, the four top final states can also fill a potential sensitivity gap in $s$-channel gluon fusion production of top-philic states $gg\to t\bar t$ that results from destructive signal-background interference, distorting or removing the resonance structure. The dominant partonic channels of four top final states show no such behaviour and therefore remain robust processes even when interference removes sensitivity in top pair gluon fusion production. In particular, this can add complementary sensitivity to BSM searches for (but not limited to) two Higgs doublet models as we have shown in a representative parameter scan for the 2HDM type II.

%%%%%%%%%%%%%%%%%%%%%%%%%%%%%%%%%%
\subsection*{Acknowledgements}
A. and this work is funded by the Leverhulme Trust under grant RPG-2021-031.
O.A. is supported by the UK Science and Technology Facilities Council (STFC) under grant ST/V506692/1.
 A.B. and C.E. are supported by the STFC under grant ST/T000945/1. C.E. acknowledges further support by the IPPP Associateship Scheme. 
W.N. is funded by a University of Glasgow College of Science and Engineering Scholarship
P.S. is supported by the Deutsche Forschungsgemeinschaft under Germany's Excellence strategy EXC2121 ``Quantum Universe" - 390833306. 
This work has been partially funded by the Deutsche Forschungsgemeinschaft (DFG, German Research Foundation) - 491245950. 

%%%%%%%%%%%%%%%%%%%
%\bibliographystyle{JHEP}
\bibliography{paper.bbl}
%%%%%%%%%%%%%%%%%%%

\end{document}